\documentclass[aps,a4paper, preprint, superscriptaddress,preprintnumbers,floatfix,nofootinbib,amsmath,amssymb]{revtex4}
\usepackage{hyperref}
\usepackage{tikz}   
\usepackage{tikz-feynman}
\usepackage{cleveref}
\usepackage{subfig}

\usepackage{amstext,amssymb}
\usepackage[section]{placeins}
\usepackage{amsmath}
\usepackage{graphicx}
\usepackage{xspace}
\usepackage{color}
\usepackage{float}
\usepackage{comment}
\usepackage{amstext,amssymb}
\usepackage{amsmath}
\usepackage{graphicx}
\usepackage[section]{placeins}
\usepackage{units}
\usepackage{array}
\usepackage{amsmath,bm}
\usepackage{amssymb}
\usepackage{gensymb}
\usepackage{textcomp}
\usepackage{slashed} 
\usepackage{multirow}
\usepackage{hyperref}
\usepackage{mathrsfs}
\usepackage{enumitem}
\usepackage{graphicx}
\usepackage{color}
\usepackage{comment}
\usepackage{epstopdf}
\usepackage{float}
\usepackage{units}
\usepackage{soul}
\usepackage{feynmp}
\usepackage{slashed}
\definecolor{antiquewhite}{rgb}{0.98, 0.92, 0.84}
\definecolor{aqua}{rgb}{0.0, 1.0, 1.0}
\definecolor{amethyst}{rgb}{0.6, 0.4, 0.8}
\definecolor{applegreen}{rgb}{0.55, 0.71, 0.0}
\definecolor{byzantine}{rgb}{0.74, 0.2, 0.64}
\definecolor{cadetgrey}{rgb}{0.57, 0.64, 0.69}
\definecolor{candypink}{rgb}{0.89, 0.44, 0.48}
\usepackage{amsmath,bm}
\usepackage{slashed} 
\usepackage{multirow}
\newcommand{\be}{\begin{equation}}
\newcommand{\ee}{\end{equation}}
\newcommand{\bea}{\begin{eqnarray}}
\newcommand{\eea}{\end{eqnarray}}
\newcommand{\nn}{\nonumber}

\def\s1{\hat s}

\usepackage{slashed}

\newcommand{\nua}[1]{\ensuremath{\rlap{\kern-2.5pt\ensuremath{\overset{\scriptscriptstyle(-)}{\phantom{\nu}}}}{\ensuremath{{\nu}_{#1}}}}\xspace}

\begin{document}
\title{Unveiling neutrino phenomenology, $(g-2)_{e,\mu}$ and leptogenesis through U(1) gauge symmetries in an inverse seesaw model}
\author{Papia Panda}
\email{ppapia93@gmail.com}
\affiliation{School of Physics,  University of Hyderabad, Hyderabad - 500046,  India}
\author{Mitesh Kumar Behera}
\email{miteshbehera1304@gmail.com}
\affiliation{School of Physics,  University of Hyderabad, Hyderabad - 500046,  India}
\author{Priya Mishra }
\email{mishpriya99@gmail.com}
\affiliation{School of Physics,  University of Hyderabad, Hyderabad - 500046,  India}
\author{ Rukmani Mohanta}
\email{rmsp@uohyd.ac.in}
\affiliation{School of Physics,  University of Hyderabad, Hyderabad - 500046,  India}

\begin{abstract}
The proposed work is an extension of the Standard Model, where we have introduced two gauge symmetries, i.e., $U(1)_{B-L}$ and $U(1)_{L_e-L_\mu}$ to study neutrino phenomenology, muon, and electron $(g-2)$ as well as leptogenesis using the inverse seesaw mechanism. For this purpose, we have included three right-handed neutrinos $N_{R_i}$, three neutral fermions $S_{L_i} (i =1,2,3)$ and two scalar singlet bosons ($\chi_1$ and $\chi_2$). We get a definite structure for the neutrino mass matrix due to the aforementioned gauge symmetries. Thus, our model is able to predict the neutrino oscillation results, which are in accordance with the experimental data and is inclined towards normal ordering. The outcomes comprise the active neutrino masses, mixing angles, mass squared differences, CP-violating phase, etc. 
Moreover, since the extended gauge symmetries are local, there are corresponding gauge bosons, denoted as $Z_{B-L}$ and  $Z_{e \mu}$. Of these, mass of $Z_{B-L}$ is  $\mathcal{O}$(TeV) range to satisfy the collider constraint, while the mass of $Z_{e \mu}$ is in the MeV range, making it feasible to account for current electron and muon $(g-2)$ results via neutral current interactions. Furthermore, our model is able to account for leptogenesis, which can demonstrate  the matter-antimatter asymmetry of the universe. Additionally, we have carried out the prospect  of probing our model in the context of upcoming long baseline experiments: DUNE, T2HK, and T2HKK, at a confidence level of $5\sigma$. From the result it is clear that, our model can be tested in its $3\sigma$ C.L. with $5\sigma$ allowed region of DUNE, T2HK and T2HKK.
\end{abstract}

\maketitle
\flushbottom

\section{Introduction }
\label{introduction}
The standard model (SM) is inordinately successful in explaining most of the properties of the hadrons, charged leptons, and gauge bosons with high precision. In spite of exceptional triumph, it lacks in explaining the tininess  of neutrino mass \cite{Ma:1998dn}, nature of the neutrinos, i.e., Dirac or Majorana,  matter-antimatter asymmetry, muon and electron anomalous magnetic moments $(g-2)_{\mu, e}$, dark matter content of the  universe, etc. 
To work within the domain of SM and generate small neutrino mass, the Weinberg operator \cite{Weinberg:1979sa} becomes an unavoidable requisite. However, sketching a model feasible for explaining other puzzling phenomena along with neutrino phenomenology seems difficult via the dimension-5 operator. Therefore, to leap beyond the standard model (BSM) becomes more of an obligation, and to do so, we introduce right-handed neutrinos (RHNs) into the picture, which lays the foundation for the seesaw mechanism. Type-I or canonical \cite{Ma:2005py,Palcu:2009uk,Zhang:2011vh} is the simplest seesaw mechanism which utilizes only three additional SM singlet RHNs to showcase the smallness of neutrino mass. The neutrino mass formula in this framework takes the form shown below
\begin{equation}
m_\nu= - \mathcal{M}_D \mathcal{M}_R^{-1} \mathcal{M}^T_D\;,
\label{type1-nmass}
\end{equation}   
where $\mathcal{M}_D, \mathcal{M}_R $ are the Dirac and heavy RHN mass matrices, respectively. In order to bring the scale of active neutrinos in the sub-eV range as constrained by cosmological bound \cite{DiValentino:2019dzu}, one needs heavy RHNs to be as massive as ${\cal O}( 10^{14}) $ GeV. To prove the existence of such a massive neutrino is impractical for current experiments. Hence, its usage is very prevalent, as seen in myriad literature. On the other hand, additional variants of the seesaw are: type-II seesaw with scalar triplets \cite{Rodejohann:2004cg,Ding:2017jdr,deSousaPires:2018fnl,Dong:2010gk}, type-III seesaw with fermion triplets \cite{Franceschini:2008pz,Goswami:2018jar,Biswas:2019ygr,FileviezPerez:2008sr,Mishra:2023cjc,Mishra:2022egy}, linear seesaw \cite{Arbelaez:2021chf,Behera:2020sfe,Behera:2022wco,Hirsch:2009mx,Deppisch:2015cua}, inverse seesaw \cite{Ma:2009gu,Behera:2020lpd,Behera:2021eut,Parida:2010wq,Dias:2012xp,Dias:2011sq,CentellesChulia:2020dfh,Pinheiro:2021mps,Pongkitivanichkul:2019cvm,Abada:2021yot,Nomura:2019xsb,An:2011uq,Biswas:2018yus} etc. In this article, we investigate the inverse seesaw scenario utilizing extended symmetries, which involves $U(1)_{B-L}$ and $U(1)_{L_e-L_\mu}$ along with three right-handed neutrinos $N_{R_i}$ and three neutral fermions $S_{L_i}$, where $i = 1,2,3$. By incorporating these six heavy neutrinos, we can maintain the inverse seesaw mass matrix structure described in eqn. (\ref{eq:numatrix-complete}) as long as appropriate charge assignments are made under the extended symmetries. 
In the present work, due to the presence of $U(1)_{L_e-L_\mu}$ gauge symmetry, additional one-loop contributions to electron and muon $(g-2)$ through the new gauge boson $Z_{e \mu}$ are  possible. Thus, our model is able to explain both electron and muon  anomalous magnetic moments $(g-2)_{e,\mu}$. Additionally, in this work, we have shown the testability of our model in three future long-baseline experiments, DUNE, T2HK, and T2HKK.

The primary motivation for utilizing two anomaly-free gauge symmetries in conjunction with the SM symmetries is twofold: the $U(1)_{B-L}$ symmetry allows us to generate small active neutrino mass, facilitate leptogenesis (discussed in sec. \ref{sec:lepto}), and provide collider search opportunities. Meanwhile, the MeV range gauge boson arising from the $U(1)_{L_e-L_\mu}$ gauge symmetry aids in explaining the anomalous magnetic moments of both electrons and muons. An attractive aspect of these symmetries is that one of them ($U(1)_{L_e-L_{\mu}}$) is anomaly-free, and another one ($U(1)_{B-L})$ becomes anomaly free only by adding three right-handed neutrinos with $B-L$ value of $(-1)$ each, thus enabling us to account for various aspects of neutrino phenomenology, such as the measured values of mixing angles ($\sin^2 \theta_{13},~ \sin^2 \theta_{12} $ and $\sin^2 \theta_{23}$), mass-squared differences ($\Delta m^2_{31}$ and $\Delta m^2_{21}$), CP-violating phase ($\delta_{\rm CP}$), etc. We also discuss the intriguing results of electron and muon $(g-2)$. Previous studies that utilized the $U(1)_{L_\mu - L_\tau}$ symmetry to explain both neutrino phenomenology and muon $(g-2)$ have been successful but lack direct experimental evidence. The gauge boson resulting from the symmetry breaking of $U(1)_{L_\mu - L_\tau}$ would not be directly observable in the current LHC or the $e^-e^+$ machines. On the other hand, the gauge boson associated with $U(1)_{B-L}$ can be detected by LHC experiments, and $e^- e^+$ collider experiments can determine the upper limit of the mass of the gauge boson related to $U(1)_{L_e-L_\mu}$. These observations motivate us for the use of $U(1)_{B-L} \times U(1)_{L_e - L_\mu}$ as an extended gauge symmetry. As mentioned earlier, local gauge symmetries require the cancellation of gauge anomalies, which can be achieved by assigning appropriate charges to the $N_{R_i}$ and $S_{L_i}$ under the extended symmetries. Additionally, two singlet scalar bosons $(\chi_1$ and $\chi_2)$ are introduced to contribute to symmetry breaking, resulting in the existence of associated gauge bosons $Z_{B-L}$ and  $ Z_{e \mu}$.

This paper provides a detailed outline of the model framework, including particle content, Lagrangian, and mass matrices in Section \ref{model description}. The article then goes on to present numerical calculations in Section \ref{numerical calculation}, offering values for the allowed parameter space that helps explain the behavior of neutrino oscillation data. It's crucial to test this proposed model in upcoming neutrino experiments; hence section \ref{sec:long_baseline} provides a comparative study of the testability of the proposed model in various future long baseline experiments, such as DUNE, T2HK, and T2HKK, which offer great precision toward the measurement of neutrino oscillation parameters.  Electron and muon $(g-2)$ explanations are covered in Section \ref{electon-and-muon-g-2}, while Section \ref{sec:lepto} focuses on leptogenesis with possible benchmark points to ensure the correct baryon asymmetry of the Universe. In Section \ref{coll-bound}, a brief discussion  on the collider search  for the gauge boson $Z_{B-L}$ associated with $U(1)_{B-L}$ is presented. Finally, the results of this study are summarized in Section \ref{sec: concl}.

\section{Model Framework }
\label{model description}
\begin{table}[htb]
\begin{center}
\begin{tabular}{|c|c|c|c|c|c|}
	\hline
			Particles &  $SU(3)_C $   &   $SU(2)_L$ & $ U(1)_Y$	& $U(1)_{B-L}$	& $U(1)_{L_{e}-L_{\mu}}$\\
	\hline
	$\ell^{}_{{\rm {\alpha L}}} ( \alpha= e, \mu, \tau)$	& $\textbf{1} $ & $\textbf{2}$	&  $-1$	& $-1$   &  $1,-1,0$\\
			 $\ell^{}_{\rm {\alpha R}} (\alpha = e, \mu, \tau)$		& $\textbf{1} $					& $\textbf{1}$	&  $-2$	& $-1$    &   $1,-1,0$   \\
			 $N^{}_{{\rm R_i}} (i = 1, 2, 3)$			& $\textbf{1} $			& $\textbf{1}$	&  $0$	& $-1$    &   $1,-1,0$   \\
	$S^{}_{{\rm L_i}} ( i = 1, 2, 3)$			& $\textbf{1} $			& $\textbf{1}$	&  $0$	& $0$    &   $1,-1,0$ \\		 
	\hline
	  $H$	& $\textbf{1} $						& $\textbf{2}$	&   $1$	& $0$   & $0$\\
		 $\chi_1$		& $\textbf{1} $					& $\textbf{1}$	&   $0$	& $1$    &     $0$   \\
			 $\chi_2 $		& $\textbf{1} $				& $\textbf{1}$	&  $0$	& $0$   &   $1$\\  
	\hline
\end{tabular}
\caption{Particles and their corresponding charge assignment under  $SU(3)_C \times SU(2)_L \times U(1)_Y \times U(1)_{B-L} \times U(1)_{L_e-L_\mu}$ model.}
\label{Table: model}
\end{center}
\end{table} 
As an extension of the SM, we have added three right-handed neutrinos (RHNs) $N_{R_i}$ and three neutral fermions $S_{L_i}, (i=1,2,3$) to achieve an inverse seesaw mechanism for giving mass to active neutrinos in the sub-eV scale. The charges assigned to the RHNs and neutral fermions under $U(1)_{B-L}$ is $-1$ and $0$, while, $\{1,-1,0\}$ and $\{1,-1,0\}$ under $U(1)_{L_e -L_\mu}$ respectively. In addition to the above, we have included two extra singlet scalars $\chi_1$ and $ \chi_2$ with assigned charges under $U(1)_{B-L}$ as $1, 0$ and under $U(1)_{L_e -L_\mu}$: $0,~1$ respectively.  $U(1)_{B-L}$ gauge symmetry is spontaneously broken by the vacuum expectation value (VEV) of $\chi_1$, whereas VEV of $\chi_2$ is responsible for the breaking of $U(1)_{L_e-L_\mu}$ symmetry.  
The Lagrangian for the leptonic sector of the proposed model is
\begin{eqnarray}
    \mathcal{L}_{\rm lepton}  &\supset &    \mathcal{L}_{SM}^{\rm lepton} + \left[y_D^e \bar{\ell}^{}_{e{\rm L}} \tilde{H} N_{R_1} +  y_D^\mu \bar{\ell}^{}_{\mu{\rm L}} \tilde{H} N_{R_2} 
    +  y_D^\tau \bar{\ell}^{}_{\tau{\rm L}} \tilde{H} N_{R_3}\right]\nonumber\\
    &+& \left [ y^1_{N} \bar{S}_{L_1} N_{R_1} \chi_1 +   
    y^{2}_{N} \bar{S}_{L_2} N_{R_2} \chi_1 + y^
    {3}_{N} \bar{S}_{L_3} N_{R_3} \chi_1 \right] \nonumber \\
    &+& \left[  \mathcal{M}_{1 2} \bar{S}_{L_1}^c S_{L_2} + y_{1 3} \bar{S}_{L_1}^c S_{L_3} \chi_2^* + y_{2 3} \bar{S}_{L_2}^c S_{L_3} \chi_{2} +\mathcal{M}_{3 3} \bar{S}_{L_3}^c S_{L_3}  \right] + h.c., 
    \label{eq:Yukawalep_le-ltau}
    \end{eqnarray}
 where $\mathcal{L}_{SM}^{\rm lepton} $ is the standard model Lagrangian for the leptonic sector, the three terms within bracket of the first line are the Dirac type Yukawa interaction terms between heavy right-handed neutrino and SM particles, the second line represents the mixing terms involving $N_{R_i}$ and $S_{L_i}$, and finally, the terms of third line generate the mass matrix $(\mathcal{M}_\mu$) for neutral fermions, $S_{L_i}$. 
Thus, one can obtain the  Dirac mass term $\mathcal{M}_D$, the mixing term $\mathcal{M}_{NS}$, and $\mathcal{M}_\mu$ as given below in eqs. (\ref{mass_dirac}) and (\ref{eq:MD ML_le-ltau}) 
 \begin{eqnarray}
\quad \mathcal{M}_D = \frac{v_H}{\sqrt{2}}\begin{pmatrix}
 |y_D^e| e^{i \phi_1}  &  0  &  0 \cr 0  & y_D^{\mu} & 0 \cr 0 & 0 & y_D^{\tau} \end{pmatrix} ,
 \label{mass_dirac}
\end{eqnarray}   
  
\begin{eqnarray}
\quad \mathcal{M}^{}_{ NS} =  \frac{v_1}{\sqrt{2}} \begin{pmatrix}|y^1_{N}| e^{i \phi_2} ~ &~ 0 ~&~ 0 \cr 0  ~&~ y_{N}^{2} ~&~ 0 \cr 0 ~&~ 0 ~&~  y_{N}^{3}  \end{pmatrix}  \; ,    
   \quad
     \mathcal{M}_\mu =  \left( \begin{matrix} 0 & \mathcal{M}_{1 2} & y_{1 3} \frac{v_2}{\sqrt{2}}  \cr  \mathcal{M}_{1 2} &  0 &  y_{2 3} \frac{v_2}{\sqrt{2}}  \cr  y_{1 3} \frac{v_2}{\sqrt{2}}  & y_{2 3} \frac{v_2}{\sqrt{2}}  & \mathcal{M}_{3 3}  \end{matrix} \right) \; ,
     \label{eq:MD ML_le-ltau}
\end{eqnarray}
where the VEVs of the scalars are represented as $\langle H \rangle=\displaystyle{\left (0,\frac{v_H}{ \sqrt 2}\right )^T}$,  $\displaystyle{\langle \chi_1 \rangle = \frac{v_1}{\sqrt{2}}}, ~\displaystyle{\langle \chi_2 \rangle = \frac{v_2}{\sqrt{2}}}$.
In the expression of $\mathcal{M}_D$ and $\mathcal{M}_{NS}$, we have considered $y_D^e$ and $y^{1}_{N}$ couplings to be  complex, with associated  phases $\phi_1$ and $\phi_2$, while other Yukawa couplings are real by redefinition of phases.
The neutrino mass matrix thus can have the form in the basis ($\nu_L, N_R^c, S_L$) as 
\begin{eqnarray}
\mathbb{M} = \left(\begin{array}{c|ccc}   & \nu_L & {N}_R^c  & {S}_L   \\ \hline
\nu_L  & 0    & \mathcal{M}_D       & 0 \\
{N}_R^c    & \mathcal{M}^T_D         & 0       & \mathcal{M}_{NS} \\
{S}_L & 0     & \mathcal{M}_{NS}^T    & \mathcal{M}_\mu
\end{array}
\right).
\label{eq:numatrix-complete}
\end{eqnarray}
Inverse seesaw is executed successfully by considering the assumption  $ \mathcal{M}_\mu \ll \mathcal{M}_D < \mathcal{M}_{NS}$. 
The active neutrino mass matrix $m_\nu$ can be found from the expression
\begin{equation}
m_\nu= \mathcal{M}_D^T (\mathcal{M}_{NS}^{-1})^T \mathcal{M}_\mu  \mathcal{M}_{NS}^{-1} \mathcal{M}_D\;.
\label{mnu-matrix}
\end{equation}

The kinetic terms, which contain the interactions between the scalars and gauge bosons, are given as
\begin{equation}
{\cal L}_{\rm{kin}} = |D_\mu H|^2 + |D_\mu \chi_1|^2 + |D_\mu \chi_2|^2\;,
\end{equation} 
where $D_\mu$ is the covariant derivative.  According to the quantum charges of gauge bosons, the kinetic term of the scalar sector  can be written as
\begin{eqnarray}
{\cal L}_{\rm{kin}}&=&\left |\left (\partial_\mu - i \frac{g}{2} \vec{\tau}. \vec{W_\mu} - i \frac{g^\prime}{2} {B_\mu} \right ) H\right |^2 + \left |\left (\partial_\mu - i g_{B-L} {(Z_{B-L})}_\mu \right ) \chi_1 \right |^2 \nn\\
&+& \left |\left (\partial_\mu  - i g_{e \mu} {(Z_{e \mu})}_\mu \right ) \chi_2 \right |^2\;.
\end{eqnarray}
Here the components of $\vec{\tau}$ are the Pauli matrices. $g$, $g^\prime$, $g_{B-L}$ and $g_{e \mu}$ are the gauge couplings associated to $SU(2)_L$, $U(1)_Y$, $U(1)_{B-L}$ and $U(1)_{L_e -L_\mu}$ gauge symmetries respectively. $Z_{B-L}$ and $Z_{e \mu}$ are the gauge bosons of the two additional gauge symmetries $U(1)_{B-L}$ and $U(1)_{L_e-L_\mu}$ respectively. It is worth to emphasize here that  the three $U(1)$ symmetries, i.e., $U(1)_Y$, $U(1)_{B-L}$ and $U(1)_{L_e-L_\mu}$ are spontaneously broken by the three scalar fields $H$, $\chi_1$ and $\chi_2$, when they acquire their VEVs. Each of these  scalar fields are  charged only under the $U(1)$ symmetries, which is broken by their VEVs and are neutral under the rest two symmetries. Therefore, there will be no mixing between the corresponding gauge bosons and one can safely neglect  the small  kinetic mixing between different $U(1)$ gauge bosons. As a result, the VEV of $H$ will contribute to the masses of $W^\pm$ and $Z$ bosons as
\begin{eqnarray}
 m_W^2 =\frac{g^2}{4} v_H^2,~~~{\rm and}~~~m_Z^2= \frac{1}{4}(g^2+g'^2) v_H^2\;.   
 \end{eqnarray}
 Analogously,  the mass of two new gauge bosons can be expressed as
\begin{eqnarray}
m_{{Z_{B-L}}}^2&=& \frac{1}{2}g_{B-L}^{2}v_1^2\;, \nonumber \\
m_{{Z_{e \mu}}}^2 &=& \frac{1}{2}g_{e \mu}^2 v_2^2\;.
\end{eqnarray}
With the values of VEVs (mentioned in Table \ref{table: best fit}) of two gauge bosons and their respective couplings, we can find that the mass of one gauge boson ($Z_{B-L}$) is in  TeV range while  $m_{Z_{e \mu}}$ is in MeV range.

\subsection*{Scalar sector}
\label{scalar-sector}
In this model, we have two singlet scalar bosons $\chi_1 $ and $\chi_2$ to break the extra gauge symmetries beyond the SM. The complete scalar potential of the theory is given as,
\begin{eqnarray}
 V&=&m_H^2(H^{\dagger} H)+ \lambda_1 (H^{\dagger} H)^2 + m_{1}^2(\chi_1^* \chi_1) + m_2^2 (\chi_2^* \chi_2)+ \lambda_2 (\chi_1^* \chi_1)^2 \nonumber \\
 &+&
 \lambda_3 (\chi_2^* \chi_2)^2+ \lambda_4 (H^{\dagger} H) (\chi_1^* \chi_1)+ \lambda_5 (H^{\dagger} H) (\chi_2^* \chi_2)+ \lambda_6 (\chi_1^* \chi_1)(\chi_2^* \chi_2)\;,  
 \label{potential}
\end{eqnarray}  
where $\lambda_i, (i=1,2,3,4,5,6)$ are the quartic couplings, and $m_H^2, m_{1}^2$ and $m_{2}^2  $ are the mass terms for three scalar bosons $H, \chi_1$ and $\chi_2$ respectively.

\subsection*{CP-even sector:}
The neutral component of the Higgs field  $H^0$ and the scalar fields $\chi_1$ and $\chi_2$ can be parameterized in the basis of real and pseudo scalars as
\begin{eqnarray}
H^0=\frac{v_H+ h+ i \eta}{\sqrt{2}}\;,~~~
\chi_1 = \frac{v_1 + \chi^\prime_1+ i \sigma_1}{\sqrt{2}}\;,~~~
\chi_2 = \frac{v_2+ \chi^\prime_2 + i \sigma_2}{\sqrt{2}}\;,
\label{cp-even}
\end{eqnarray}
with $h, \chi^\prime_1 $ and $ \chi^\prime_2 $ are the CP-even components of the scalar bosons, $\eta , \sigma_1$ and $\sigma_2 $ are the CP-odd components.  Thus, the CP-even mass matrix has the form in the basis of ($h, \chi^\prime_1 , \chi^\prime_2)$ as
\begin{eqnarray}
\mathcal{M}^2_E = \left(\begin{array}{c|ccc}   & h & \chi^\prime_1 & \chi^\prime_2   \\ \hline
h  &  2 \lambda_1 v_H^2   &  \lambda_4 v_H v_{1} & \lambda_5 v_H v_{2}  \\
\chi^\prime_1    & \lambda_4 v_H v_{1}  &  2 \lambda_2 v_{1 }^2  & \lambda_6 v_{1} v_{2}  \\
\chi^\prime_2 &\lambda_5 v_H v_{2}  & \lambda_6 v_{1} v_{2}  & 2\lambda_3 v_{2}^2 
\end{array}
\right).
\label{eq:me2}
\end{eqnarray}
As the scalar sector does not play any important role in the studies that we are interested in, we do not present the detailed discussion about it in this paper.

  \section{Numerical Calculations}
  \label{numerical calculation}
  
 \begin{table}
 \begin{center}
 \begin{tabular}{|c|c||c|c|}
 \hline
 Parameters  &  ranges &  Parameters & ranges \\
 \hline
 $y_D^e$   & $ [0.01,2] \times  10^{-6}$  &  $  y_{2 3}$    &  $[0.1,1] \times 10^{-7} $ \\ 
 $y_D^\mu $  &  $[0.1,2] \times 10^{-3}$   &  $ v_1 $    &  $[1,100] \times 10^{3} ~ \rm{GeV} $  \\
  $y_D^\tau$  &  $[1,5] \times 10^{-2} $ &  $v_2$   &  $[0.3,50] \times 10^{2} ~\rm{GeV} $  \\
  $y_{N}^{1}  $  &  $[1,2]$    & $\mathcal{M}_{1 2}$ &  $[1,3] ~ \rm{keV}   $  \\
  $ y_{N}^{2}  $  & $[0.1,2]  $   &   $\mathcal{M}_{3 3} $  & $ [0.1,3] ~\rm{keV} $ \\
  $ y_{N}^{3}  $  & $[0.1,2] $   &   $\phi_1$  & $ [0,2 \pi] ~\rm{rad} $ \\
  $ y_{1 3}  $  & $[0.7,7] \times 10^{-7} $   &  $\phi_2$   &  $[0,2\pi]$ rad \\
 \hline  
 \end{tabular}
 \caption{Allowed ranges of Yukawa couplings and VEVs for explaining neutrino phenomenology, electron and muon anomalous magnetic moment, and leptogenesis.}
 \label{table: best fit} 
 \end{center}
\end{table}

  The expressions for oscillation parameters related to the standard PMNS matrix $ \rm{U}$ are:
  \begin{eqnarray}
  \sin^2 \theta_{13} = |\rm{U}_{e3}|^2 ,~~~ \sin^2 \theta_{12} = \frac{|\rm{U}_{e2}|^2}{1-|\rm{U}_{e3}|^2}, ~~~ \sin^2 \theta_{23} = \frac{\rm|{U}_{\mu 3}|^2}{1-|\rm{U}_{e3}|^2}    \;,
  \end{eqnarray}
  where, $\rm{U}_{\alpha i}~ (\alpha=e, \mu, \tau,~ i=1,2,3) $ are the  elements of the PMNS  matrix $\rm{U}$.

In order to get neutrino phenomenology correlation plots, we make use of atmospheric and solar mass squared differences, i.e., ($\Delta m^2_{31}$ and $ \Delta m^2_{21} $) and  other observables of neutrino oscillations ($\theta_{12}, \theta_{23},\theta_{13}$) in their $3\sigma$ ranges from NuFIT data \cite{Esteban:2020cvm} as presented below:
  \begin{eqnarray}
  &&\sin^2 \theta_{13} = (0.02029-0.02391) ,~ \sin^2 \theta_{12}=(0.27-0.341), ~\sin^2 \theta_{23} =(0.406-0.620), \nonumber \\
&& \Delta m^2_{31} =(2.428-2.597) \times 10^{-3}~ \rm eV^2 , ~ \Delta m^2_{21} = (6.82-8.03) \times 10^{-5} ~\rm eV^2 \;.
  \end{eqnarray}

 \begin{figure}[htbp]
\begin{center}
\subfloat[]{\includegraphics[height=65mm,width=75mm]{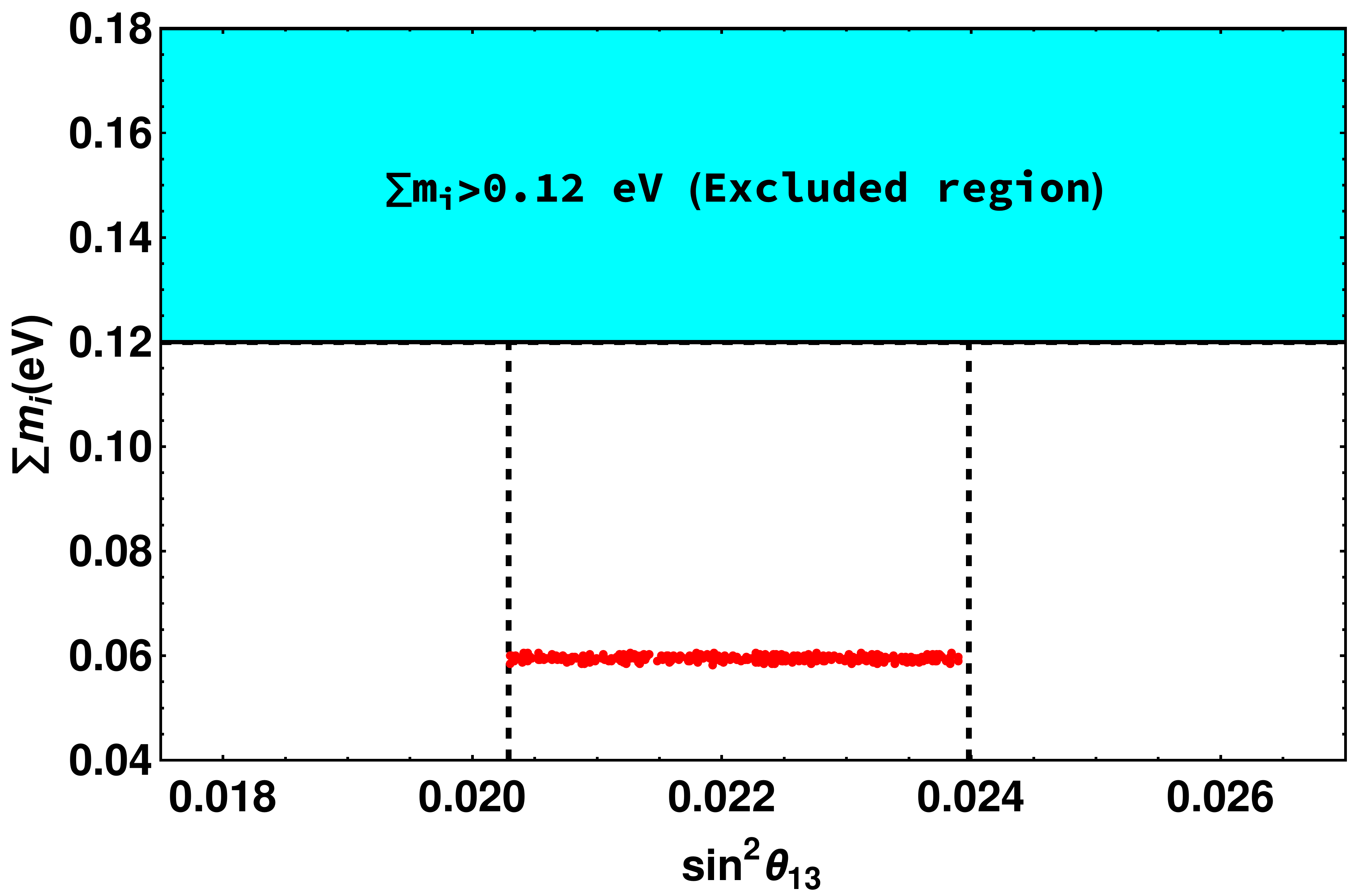}\label{nu-1}}
\hspace*{0.2 true cm}
\subfloat[]{\includegraphics[height=65mm,width=75mm]{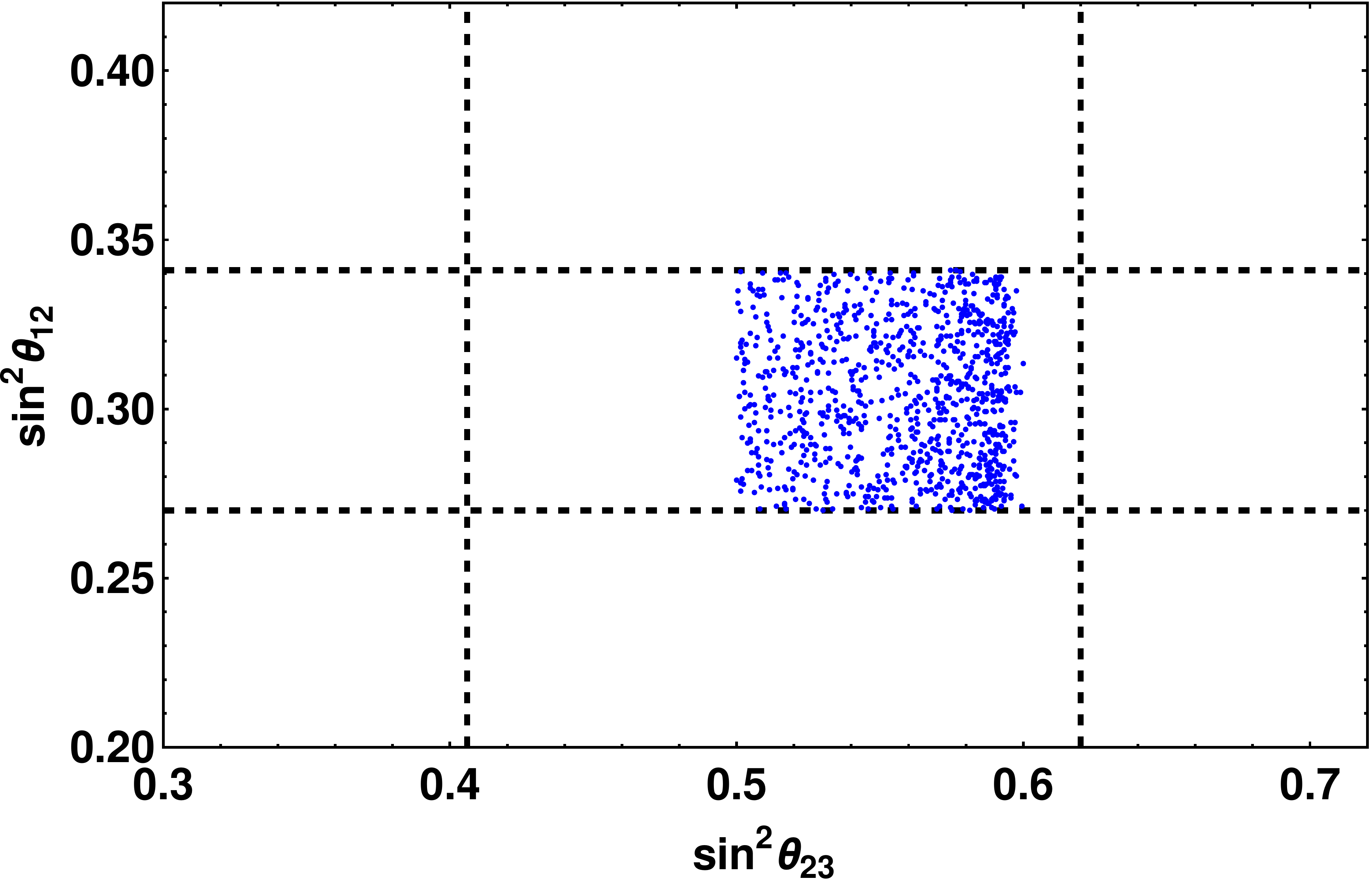}\label{nu-2}}
\caption{Panel \textbf{(\ref{nu-1})} shows the variation of active neutrino mass ($\sum m_i$) with mixing angle $\sin^2 \theta_{13} $, \textbf{(\ref{nu-2})} depicts mutually viable region for $\sin^2 \theta_{12}$ and $\sin^2 \theta_{23}$.}
\label{all-neu}
\end{center}
\end{figure}

\begin{figure}
\centering
\subfloat[]{\includegraphics[height=65mm,width=75mm]{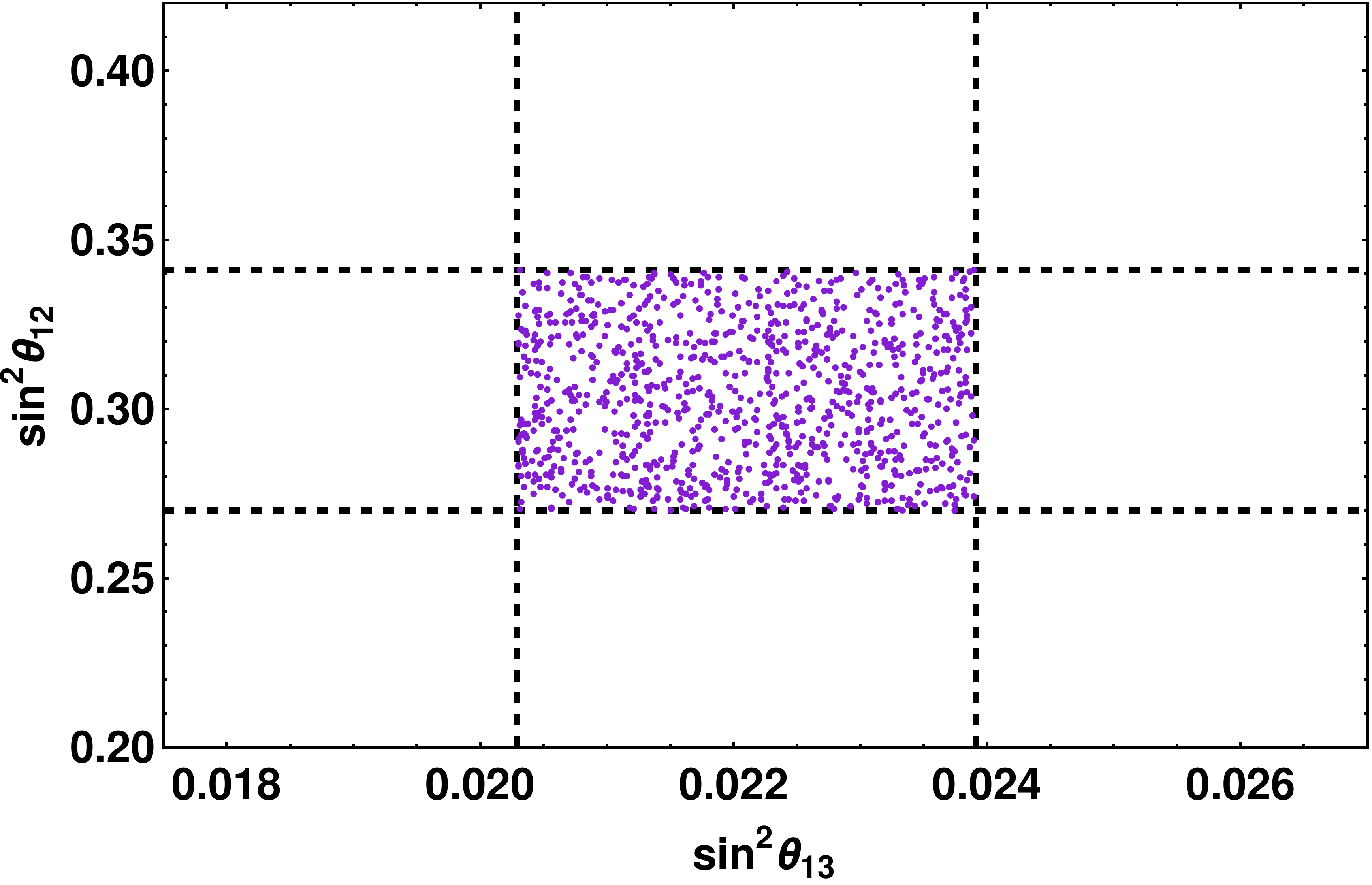}\label{nu-3}}
\hspace*{0.2 true cm}
\subfloat[]{\includegraphics[height=65mm,width=75mm]{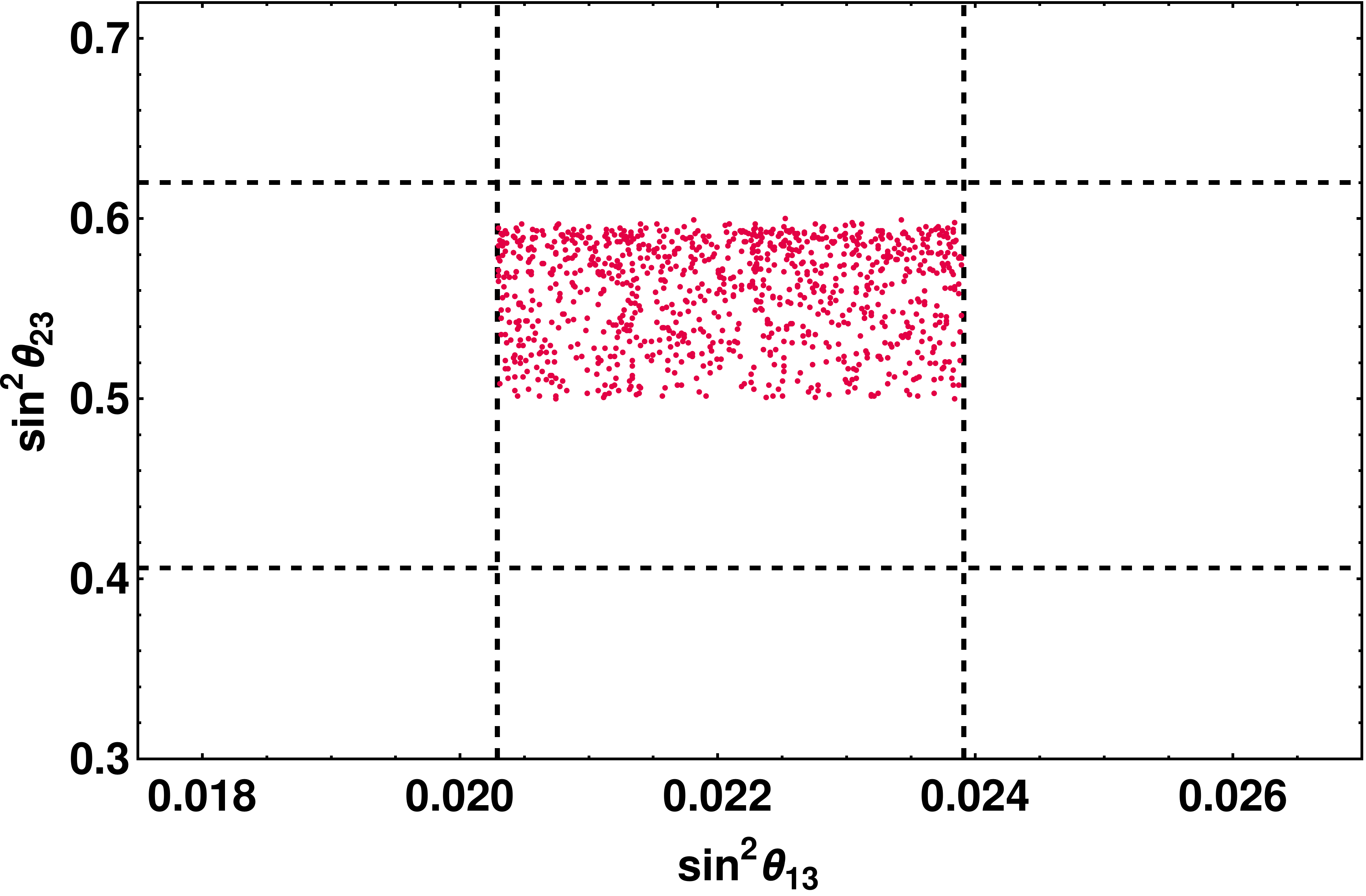}\label{nu-4}}\\
\subfloat[]{\includegraphics[height=65mm,width=75mm]{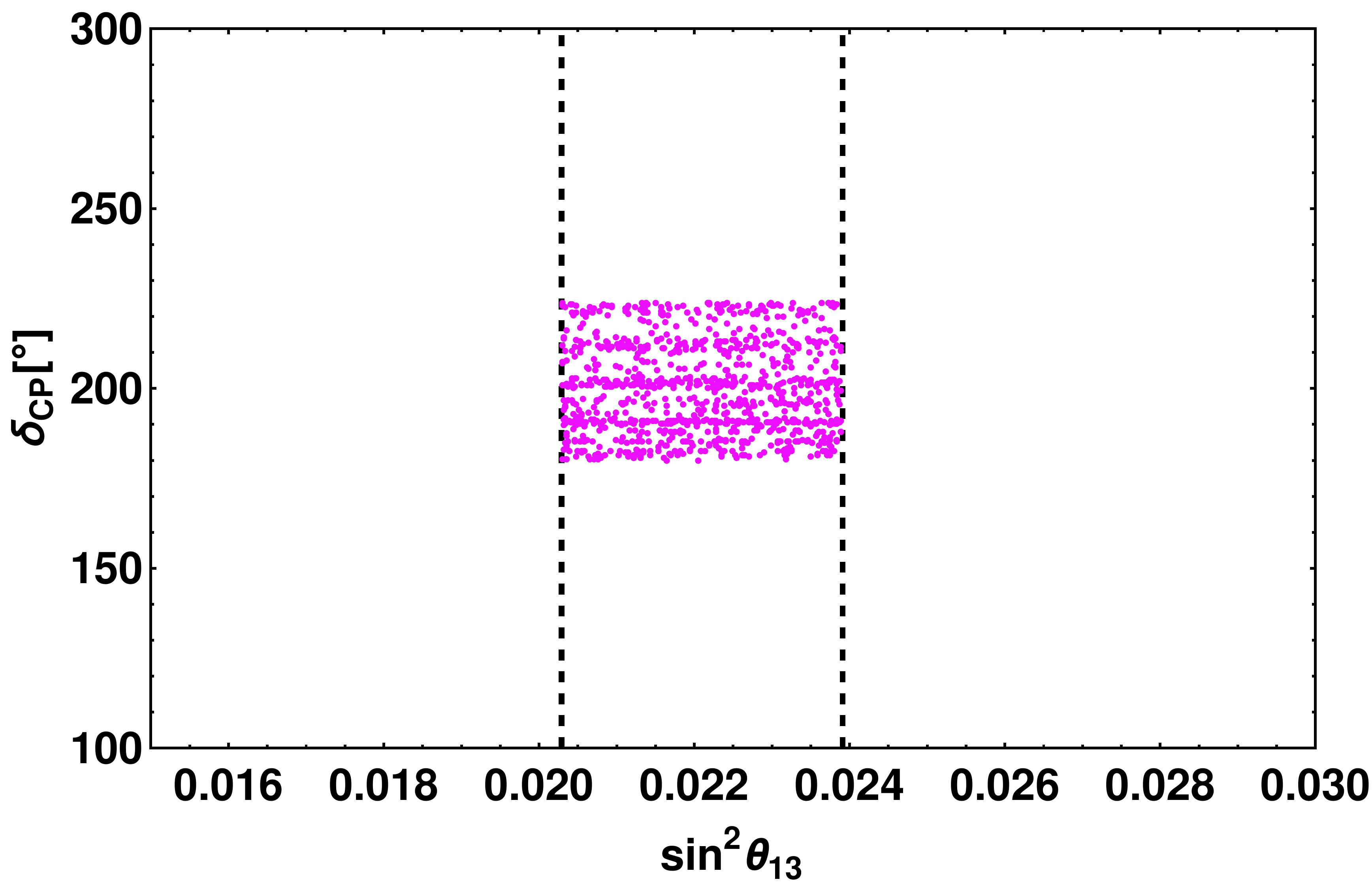}\label{nu-5}}
\hspace*{0.2 true cm}
\subfloat[]{\includegraphics[height=65mm,width=75mm]{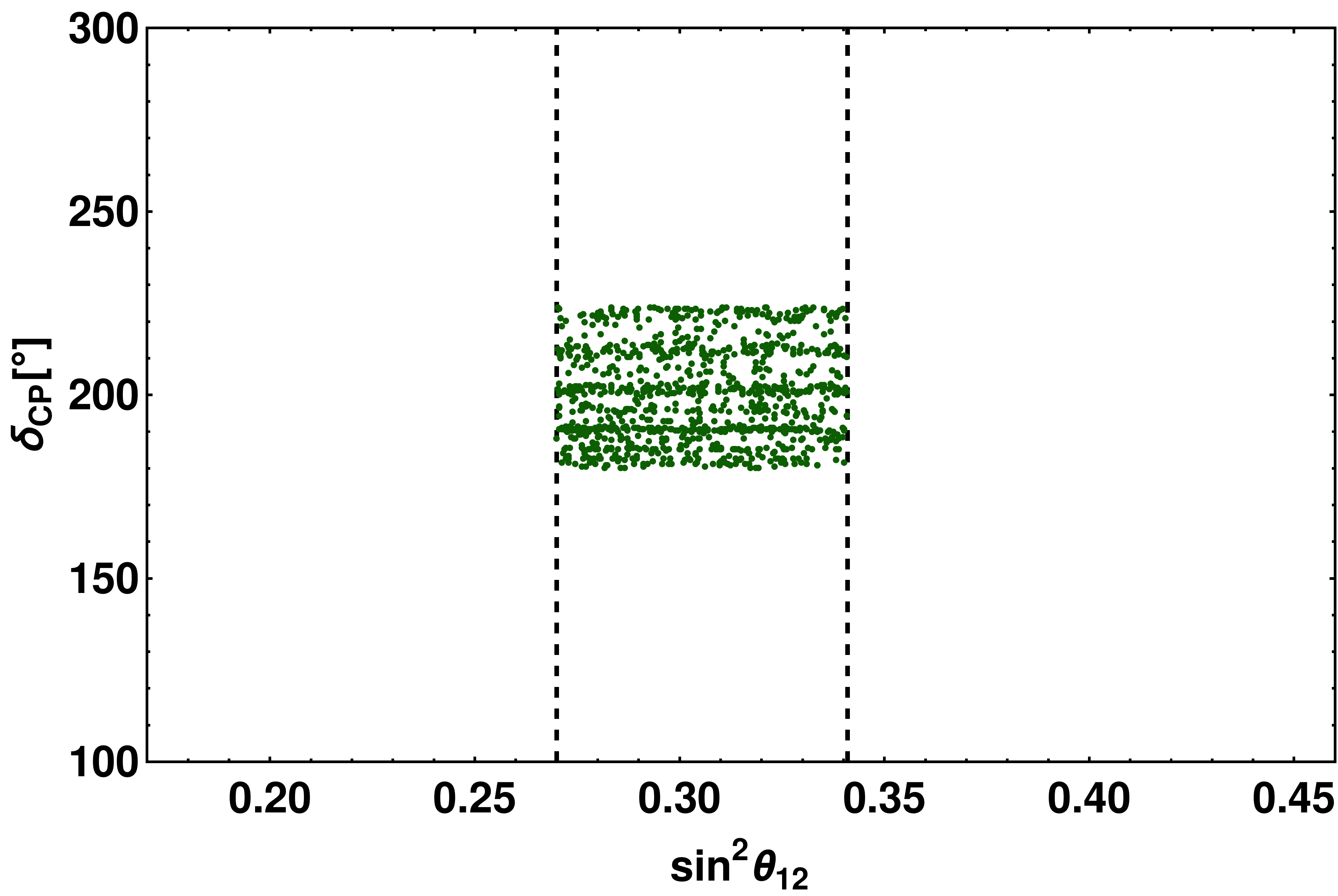}\label{nu-6}}
\vspace*{-0.2 true cm}
\caption{Panel \textbf{(\ref{nu-3})} gives the mutually allowed parameter space for mixing angle $\sin^2 \theta_{12}$ and $\sin^2 \theta_{13} $, whereas, \textbf{(\ref{nu-4})} is the correlation of $\sin^2 \theta_{13} $ with $\sin^2 \theta_{23} $, \textbf{(\ref{nu-5})} [\textbf{(\ref{nu-6})}] depicts the variation of CP violating phase $\delta_{\rm CP}$ with mixing angle $\sin^2 \theta_{13}$  [$\sin^2 \theta_{12}$].}
    \label{all-nu-2}
\end{figure}

  We have summarized the results of neutrino phenomenology in Fig.~\ref{all-neu} and \ref{all-nu-2} using ``Mixing parameter tools" \cite{Antusch:2005gp}. In Fig.~\ref{all-neu}, panel (\ref{nu-1}) shows the variation of sum of active neutrino mass 
  w.r.t. reactor mixing angle $\sin^2 \theta_{13}$. Panel (\ref{nu-2}) depicts the mutually allowed parameter space for $\sin^2 \theta_{12}$ and $\sin^2 \theta_{23}$ with black dashed gridlines indicating the corresponding $3\sigma$ allowed range. From panel~(\ref{nu-1}), we can see that, sum of active neutrino mass is very much constrained with the allowed range $0.058 ~\mathrm{eV} \leq \sum m_i \leq 0.061~ \mathrm{eV}$. From panel~(\ref{nu-2}), it is observed that $ \theta_{12}$ is unconstrained but there is a specific allowed region for $\sin^2 \theta_{23}$ ranging from $[0.5-0.6]$. Interestingly, from the panels with $\sin^2 \theta_{23}$, we can conclude that the value of $\theta_{23}$ coming from our model supports upper octant (i.e., $\sin^2 \theta_{23} > 0.5)$.  
Moving further, in Fig. \ref{all-nu-2}, panel~(\ref{nu-3}) shows the allowed parameter space for  $\sin^2 \theta_{12}$ with respect to $\sin^2 \theta_{13}$, whereas, panel~(\ref{nu-4}) gives the  correlation between  $\sin^2 \theta_{23}$ w.r.t.  $\sin^2 \theta_{13}$. From these two panels, we can observe that $\sin^2 \theta_{13}$ is fully unconstrained. In panel~(\ref{nu-5}) and panel~(\ref{nu-6}), the dependence of CP violating phase is shown w.r.t. mixing angles $\sin^2 \theta_{13}$ and $\sin^2 \theta_{12}$ respectively. These two panels describe very constrained value of $\delta_{\rm CP}$ within the range  $(180-224)^{\circ}$.

\section{Testing the model with long-baseline experiments}
\begin{figure}[htpb]
\subfloat[]{\includegraphics[height=65mm,width=75mm]{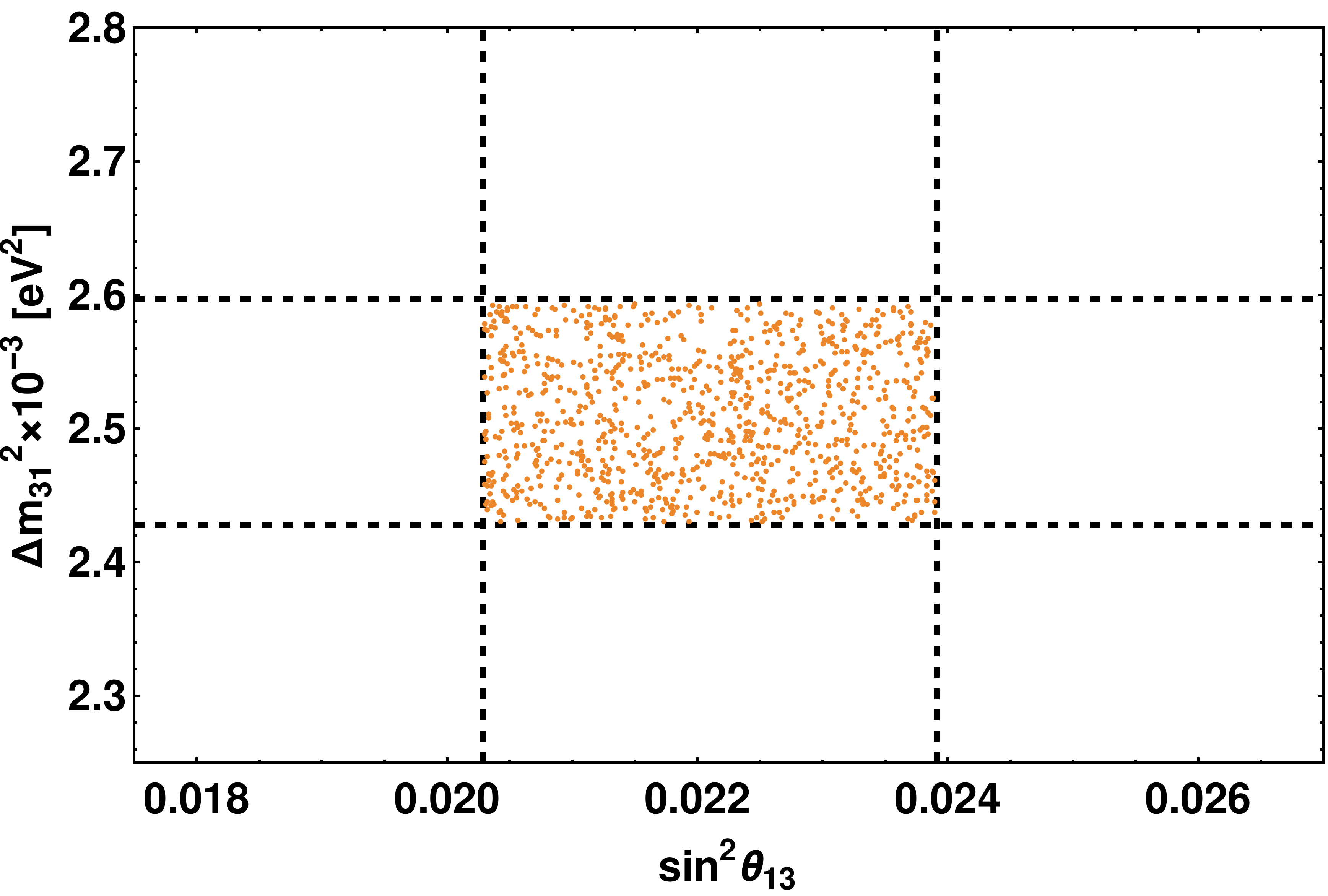}\label{ss13-dm13}}
\hspace*{0.2 true cm}
\subfloat[]{\includegraphics[height=65mm,width=75mm]{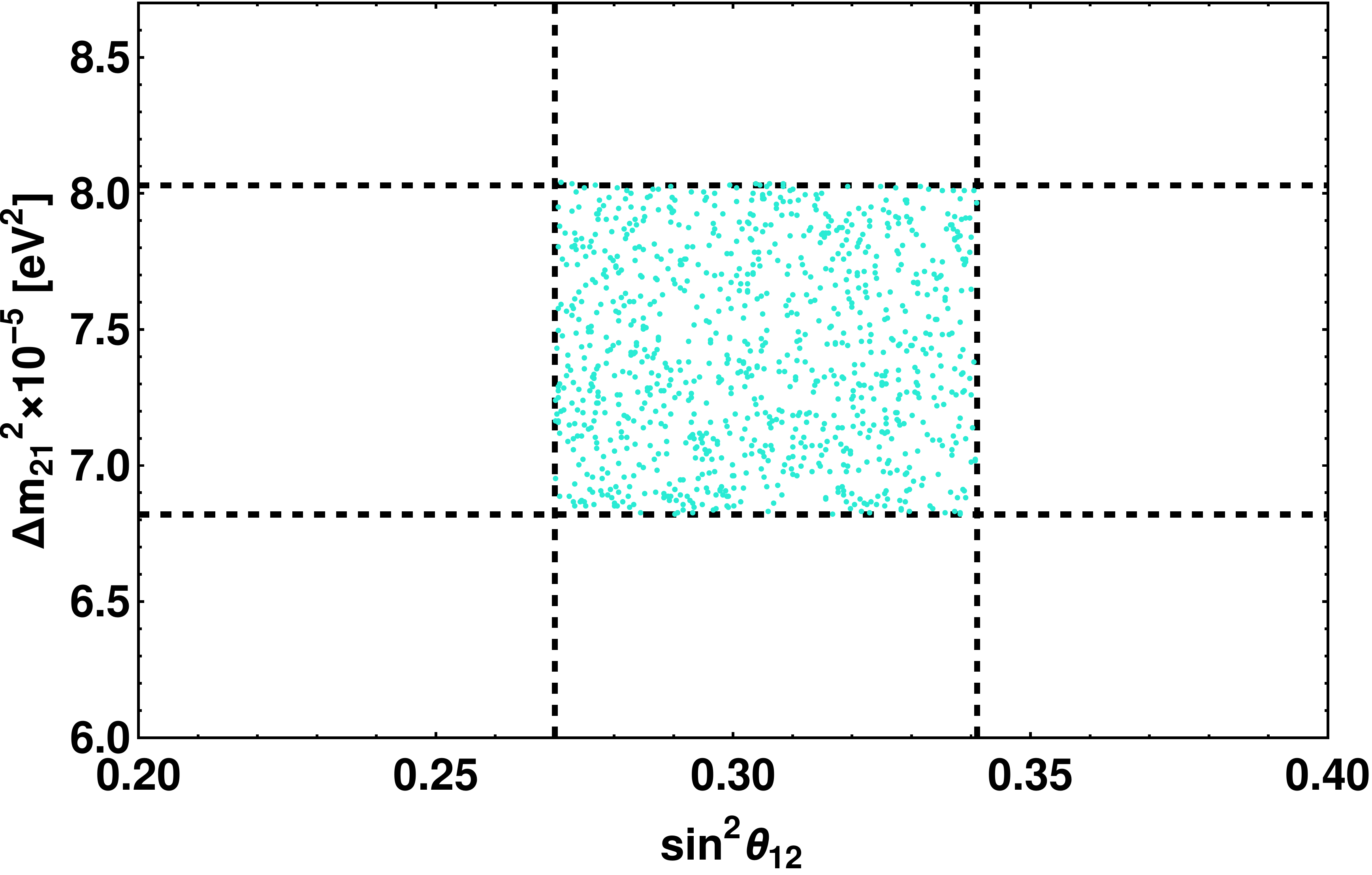}\label{ss23-dm12}}
\caption{Panel \textbf{(\ref{ss13-dm13})} shows the variation of $\Delta m_{31}^2$ w.r.t $\sin^2 \theta_{13}$, whereas, panel \textbf{(\ref{ss23-dm12})} projects the inter-dependence of $\Delta m_{21}^2$ w.r.t $\sin^2 \theta_{12}$. }
\label{nu-7}
\end{figure}
\label{sec:long_baseline}

After a detailed discussion of the model with two extended gauge symmetries ($U(1)_{B-L} \times U(1)_{L_e-L_\mu}$), in this section we try to test the proposed model in the context of future long-baseline experiments: DUNE, T2HK, and T2HKK. DUNE (Deep Underground Neutrino Experiment) is a future long-baseline experiment where the neutrino source will be located at FNAL (Fermi National Accelerator Laboratory) with a beam power of $1.2$ MW. The far detector will be placed at a distance of 1300 km at Sanford Underground Research Laboratory, South Dakoda, with liquid Argon TPC (time projection chamber) detector material of 40kt fiducial volume. In our study, we take five-year neutrino and five-year antineutrino mode running with a POT (proton on target) of $1.1 \times 10^{21}$. In this analysis, we have taken on-axis wide-band flux from source to detector. Another most promising future long baseline experiment T2HK (Tokai to Hyper Kamiokande), will be installed in Japan with two large water Cherenkov detectors of fiducial volume 187 kt each, having the neutrino source at JPARC, Tokai. The distance between JPARC to hyper-Kamiokande (HK) is around 295 km. The Neutrino beam for T2HK is 1.3 MW with a total exposure of $2.7 \times 10^{22}$. The flux from source to detector is at $2.5^{\circ}$ off-axis. In our study, we take the run-time of 10 years with an equal ratio, five years for neutrino mode and five years for antineutrino mode. Modified T2HK is also suggested, which is named as T2HKK (Tokai to Hyper Kamiokande and Korea). In T2HKK, one of the water Cherenkov detectors of HK will be shifted to Korea, 1100 km distant from the source JPARC. For T2HKK, we have used 295 km $2.5^{\circ}$ off-axis  flux and 1100 km $1.5^{\circ}$ off-axis flux. Similar to T2HK, we have used equal run time for neutrino and antineutrino, a total of ten years: five years neutrino and five years antineutrino mode.

\subsection*{Simulation details:}
For simulation purpose, we use 
GLoBES simulation package \cite{Huber:2004ka,Huber:2007ji}. For the estimation of the sensitivity, we use the Poisson
log-likelihood formula
\begin{equation}
    \chi^2_{\rm stat} = 2 \sum_{i=1}^n  \left[ N_i^{ \rm test} - N_i^{\rm true} - N_i^{\rm true} ~\rm {ln} \left(\frac{N_i^{\rm test}}{N_i^{\rm true}} \right) \right] \;,
\end{equation}
where $N_i^{\rm true}$ is the event rate from true spectrum 
and $N_i^{\rm test}$ is the event rate calculated from the test spectrum with `$i$' being the number of energy bins.
In the present work, we take $N_i^{\rm true}$ as the events corresponding to the NuFIT \cite{Esteban:2020cvm} neutrino oscillation parameters (see table \ref{best-fit-nufit}) and $N_i^{\rm test}$ being the data set coming from our model.

\subsection*{Results}
 
\begin{figure}
    \centering
    \subfloat[]{\includegraphics[height=65mm, width=75mm]{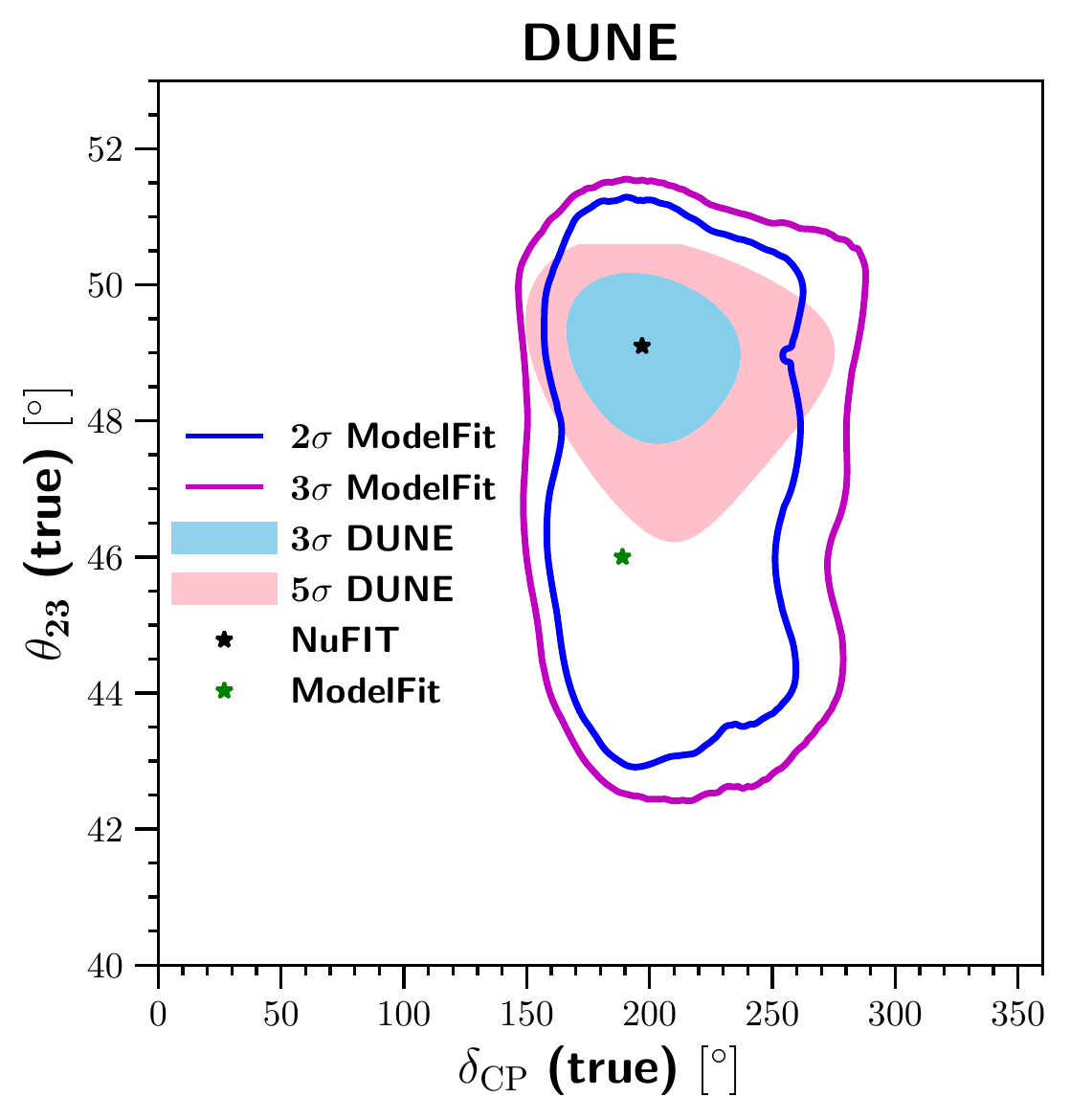}
    \label{dune-1}}
    \hspace*{0.2 true cm}
    \subfloat[]{\includegraphics[height=65mm, width=75mm]{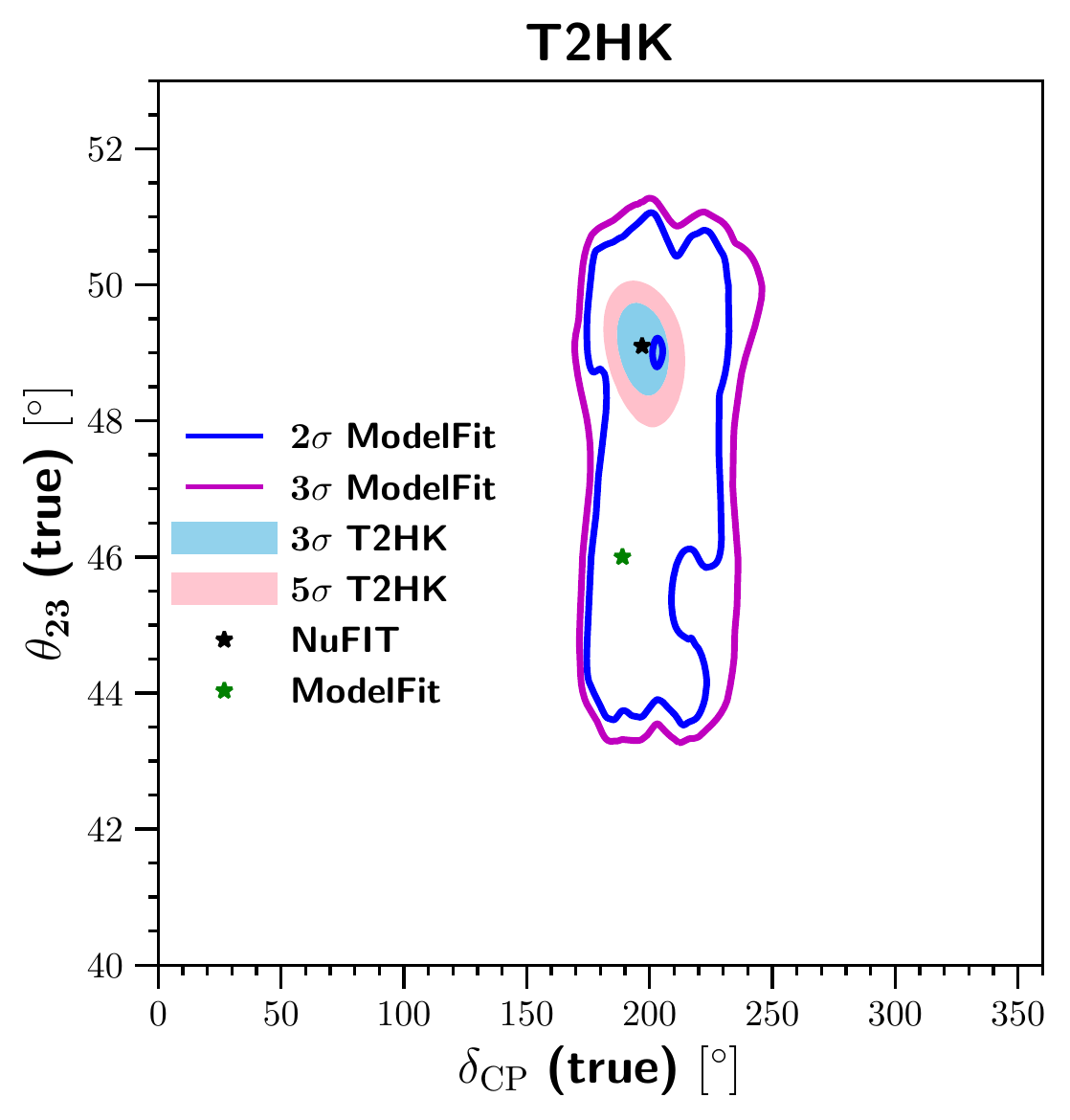}\label{t2hk}}\\
     \subfloat[]{\includegraphics[height=65mm, width=75mm]{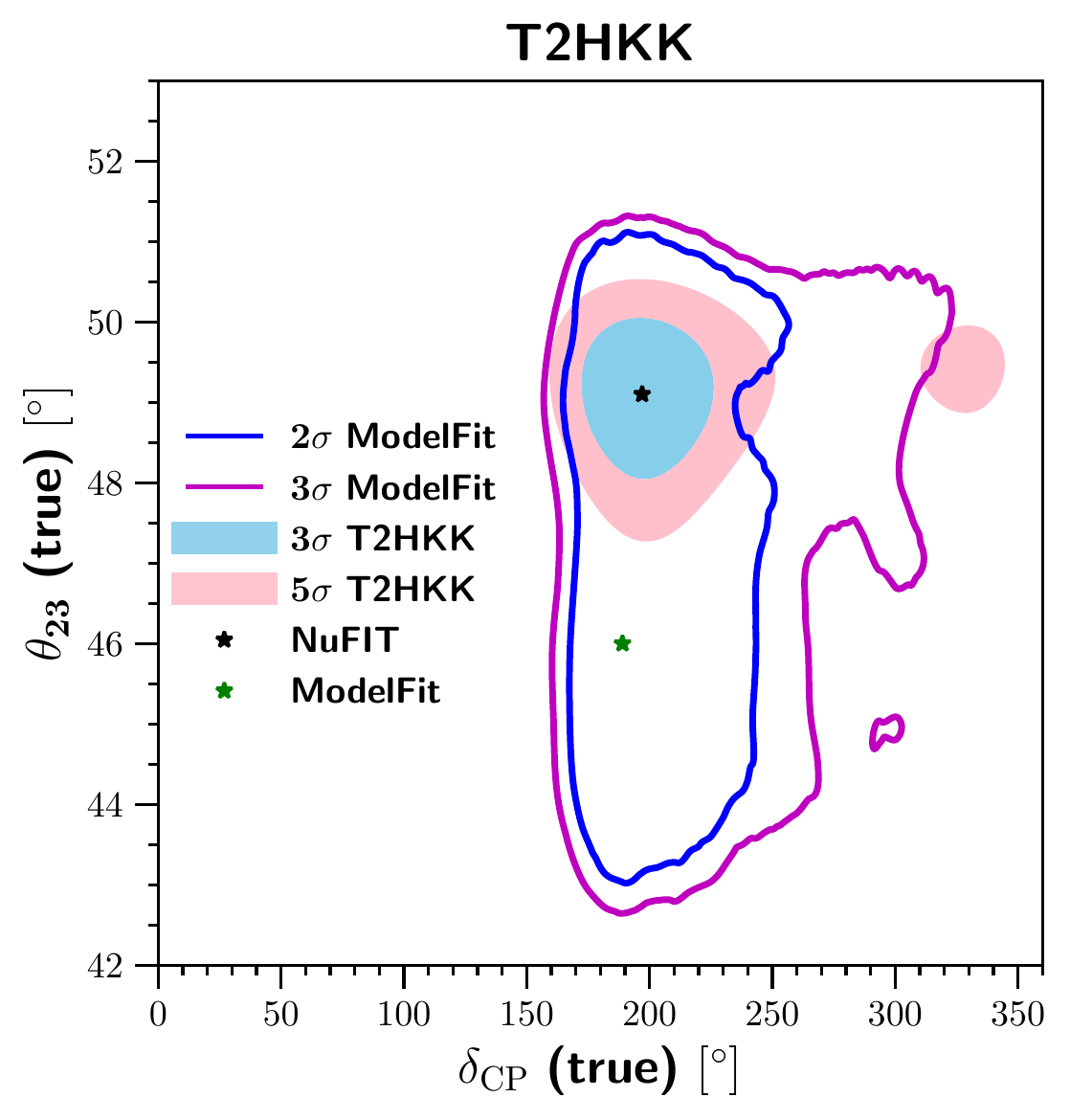}\label{t2hkk}}
 \caption{Parameter space for $\theta_{23} - \delta_{CP}$: \textbf{(\ref{dune-1})} is for DUNE experiment,
 \textbf{(\ref{t2hk})} is for T2HK and \textbf{(\ref{t2hkk})} is for T2HKK experiment. In each panel, cyan and pink shaded regions denote $3\sigma$ and $5\sigma$ allowed parameter space of corresponding experiments. Blue and magenta contours represent the $2\sigma$ and $3\sigma$ C.L. of the proposed model. The green and black stars depict the best fit values correspond to model and NuFIT respectively. }

    \label{dune-t2hk}
\end{figure}

We illustrate our results in Figs.~\ref{nu-7} and \ref{dune-t2hk}. In Fig. \ref{nu-7}, we have shown the correlation plots between neutrino oscillation parameters. 
 Panel~ \ref{ss13-dm13} [\ref{ss23-dm12}] depicts the variation of $\Delta m_{31}^2 ~
 \left[\Delta m_{21}^2\right]$ with respect to $\sin^2 \theta_{13}$ $\left[\sin^2 \theta_{12}\right]$. 
 From the panels, we can say that oscillation parameters $\Delta m_{31}^2$ and $\Delta m_{21}^2$ are unconstrained, thus our model allows all the values of $\Delta m_{31}^2$ and $\Delta m_{21}^2$ within the $3\sigma$ values of NuFIT. 
 Proceeding further, Fig. ~\ref{dune-t2hk} depicts the capability of probing our model in three future neutrino experiments, panel~(\ref{dune-1}) is for DUNE, panel ~(\ref{t2hk}) for T2HK and panel ~(\ref{t2hkk}) for T2HKK experiments respectively.

 In panel~\ref{dune-1}, cyan and pink shaded regions denote the allowed region from DUNE at $3\sigma$ and $5\sigma$ C.L. respectively by considering current best-fit values of oscillation parameters: $\theta_{23}$ and $\delta_{\rm CP}$ as the true values.
 Blue and magenta contours  represent the allowed space at $2\sigma$ and $3\sigma$ C.L. from the proposed model. Green and black stars represent the best-fit values of oscillation parameters for the model and NuFIT respectively.
 The middle column of Table \ref{best-fit-nufit} shows the best-fit values of oscillation parameters from NuFIT, whereas the rightmost column shows values of these parameters from our model corresponding to minimum $\chi^2$.
 From  panel~(\ref{dune-1}), we can see that though the best-fit values obtained from our model lie outside the $5 \sigma$ contour of DUNE,  the $2\sigma$ and $3\sigma$ allowed regions of proposed model are compatible with $5 \sigma$ parameter space of DUNE.

 Moving further, panel~(\ref{t2hk}) shows the testability of the model in the parameter space of the T2HK experiment. Analogous to DUNE, here too, cyan and pink shaded regions denote $3\sigma$ and $5\sigma$ allowed parameter space which will be covered by T2HK, if the current best-fit values of oscillation parameters remain same in future. Blue and magenta contours represent the $2\sigma$ and $3\sigma$ C.L. emerging from the proposed model.  Green and black stars represent the best-fit values of oscillation parameters for the model and NuFIT, respectively. T2HK has shorter baseline than DUNE, so statistics for T2HK is larger than DUNE. As sensitivity directly depends on statistics, the allowed parameter space of $\theta_{23}$ and $\delta_{\rm CP}$ is small for T2HK than for DUNE.
 From panel~(\ref{t2hk}), we can observe that, $3\sigma$ allowed region of our model can be tested by $5\sigma$ allowed region of T2HK. Interestingly, a small contour of parameter space which is excluded by $2\sigma$ C.L. of the model, is compatible with $3\sigma$ allowed region of T2HK. Thus, if this region is present in T2HK with the current best-fit values of NuFIT as true values, one can exclude the proposed model by $2\sigma$ C.L.

 In T2HKK experiment, one of the two water Cherenkov detectors has been shifted to Korea, 1100 km distance from JPARC. As the baseline of T2HKK is greater than T2HK, overall statistics for T2HKK is less than T2HK, resulting a wide allowed parameter space for $\theta_{23}-\delta_{\rm CP}$ plane. 
 Panel ~(\ref{t2hkk}) shows the results for T2HKK experiment. Color code for this panel is exactly same as panel ~(\ref{dune-1}) and panel ~(\ref{t2hk}). From the panel, we can conclude that $2\sigma$ allowed region is testable by $3\sigma$ allowed space of T2HKK.
 Further, the $3\sigma$ allowed region of our model can be probed  by $5\sigma$ C.L. of T2HKK.

\vspace{-0.6cm}
\begin{table}[]
    \centering
    \begin{tabular}{||c||c||c||}
    \hline \hline
      Oscillation parameters   & NuFIT  &   Model-fit \\
      \hline \hline
      $\sin^2 \theta_{13}$    & 0.02203 &  0.02110\\
      $\sin^2 \theta_{12}$     &   0.303  &  0.301\\
      $\sin^2 \theta_{23}$      &   0.572 &  0.517\\
      $\Delta m_{12}^2$      &   0.0000741 &  0.0000730\\
      $\Delta m_{23}^2$        &    0.002511  &  0.002573\\
      $\delta_{CP}$   &   $197^{\circ}$  &  $189^{\circ}$\\
      \hline \hline
    \end{tabular}
   \caption{This table showcases the data from NuFIT alongside model fit for all the oscillation parameters.}
    \label{best-fit-nufit}
\end{table}
\section{Muon and Electron \lowercase{$(g-2)$}}
\label{electon-and-muon-g-2}

\subsection{Electron $(g-2)$}

\begin{figure}
\begin{tikzpicture}
\begin{feynman}
\vertex (a);
\vertex [left = of a] (b);
\vertex[below right = of a] (c);
\vertex[above right = of c] (d);
\vertex[right = of d]  (e);
\vertex[below = of c] (f);

\diagram [line width=1.5pt]{
       (b) -- [violet,fermion, edge label=$l^-$] (a) -- [violet,fermion, edge label'=$l^-$]  (c);
       (c)   -- [violet,fermion, edge label'=$l^-$] (d);
       (c) -- [red,boson,edge label=$\gamma$] (f);
     (a) -- [blue, boson, half left, edge label=$Z_{B-L}$] (d);
     (d) --[violet,fermion, edge label=$l^-$]  (e);
};
\end{feynman}
\end{tikzpicture}
\hspace{3.5 true cm}
\begin{tikzpicture}
\begin{feynman}
\vertex (a);
\vertex [ left = of a] (b);
\vertex[below right = of a] (c);
\vertex[above right = of c] (d);
\vertex[right = of d]  (e);
\vertex[below = of c] (f);

\diagram [line width=1.5pt]{
       (b) -- [violet,fermion, edge label=$l^-$] (a) -- [violet,fermion, edge label'=$l^-$]  (c);
       (c)   -- [violet,fermion, edge label'=$l^-$] (d);
       (c) -- [red,boson,edge label=$\gamma$] (f);
     (a) -- [blue, boson, half left, edge label=$Z_{e \mu}$] (d);
     (d) --[violet,fermion, edge label=$l^-$]  (e);
};
\end{feynman}
\end{tikzpicture}
\caption{Possible Feynman diagrams for electron and muon $(g-2)$ calculation. In both the figures, $l^-=(e^-,\mu^-)$.}
\label{feyn}
\end{figure}
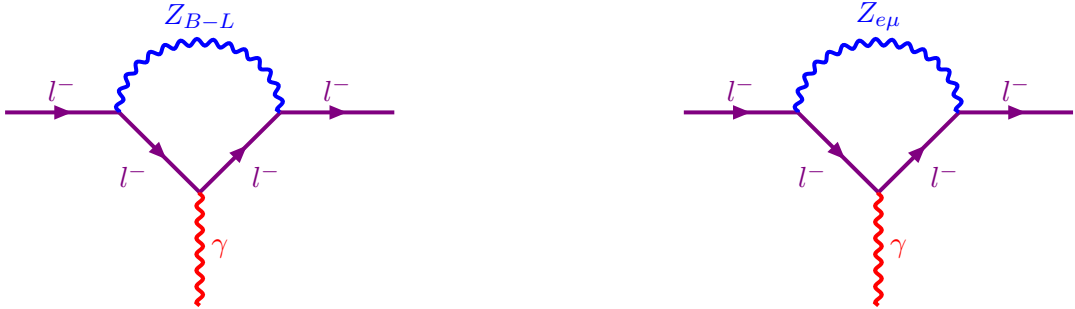

\begin{figure}[htbp]
\begin{center}
\subfloat[]{\includegraphics[height=65mm, width=75mm]{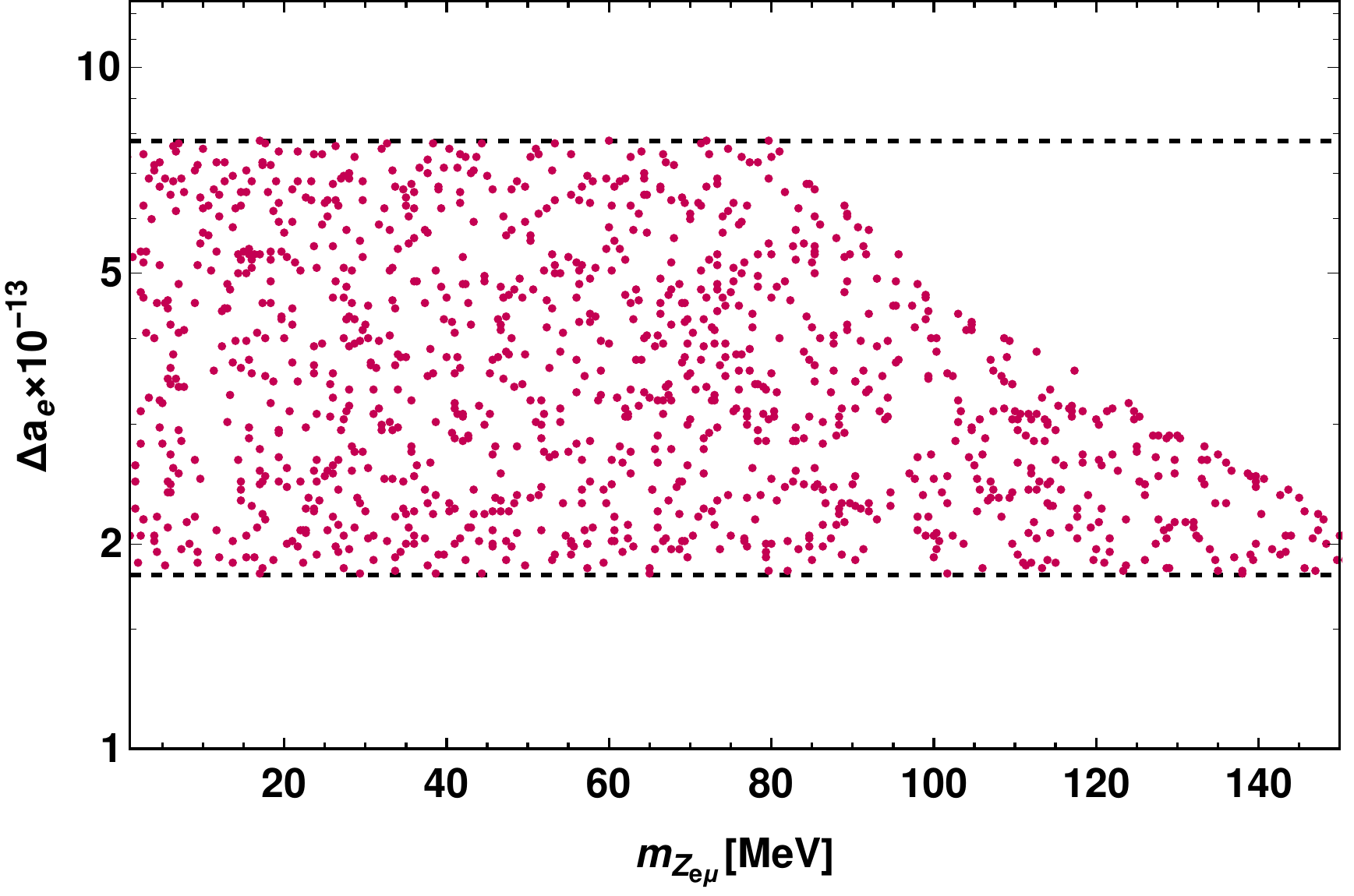}\label{amug-1}}
\hspace*{0.2 true cm}
\subfloat[]{\includegraphics[height=65mm, width=75mm]{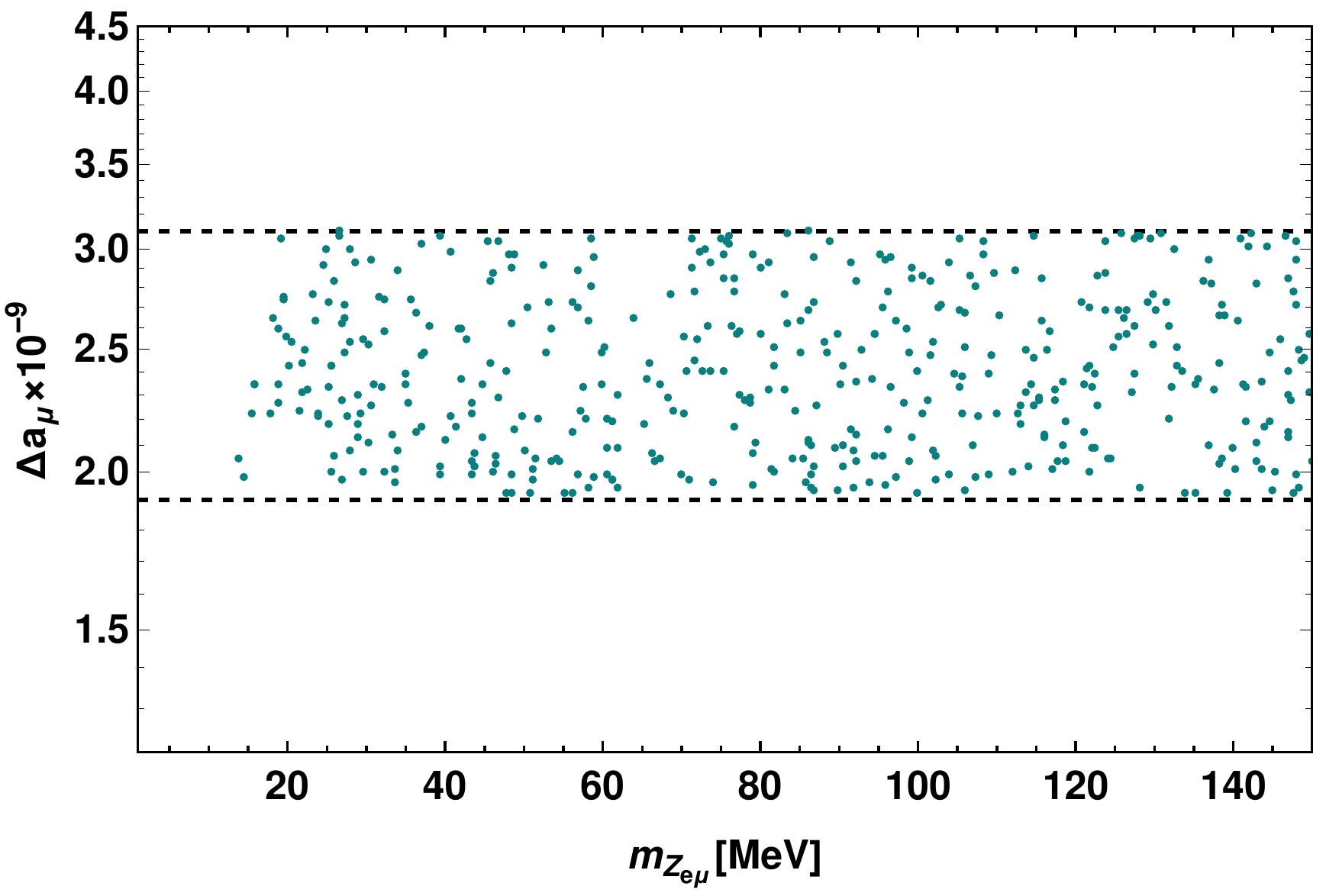}\label{amug-2}}\\
\subfloat[]{\includegraphics[height=75mm, width=95mm]{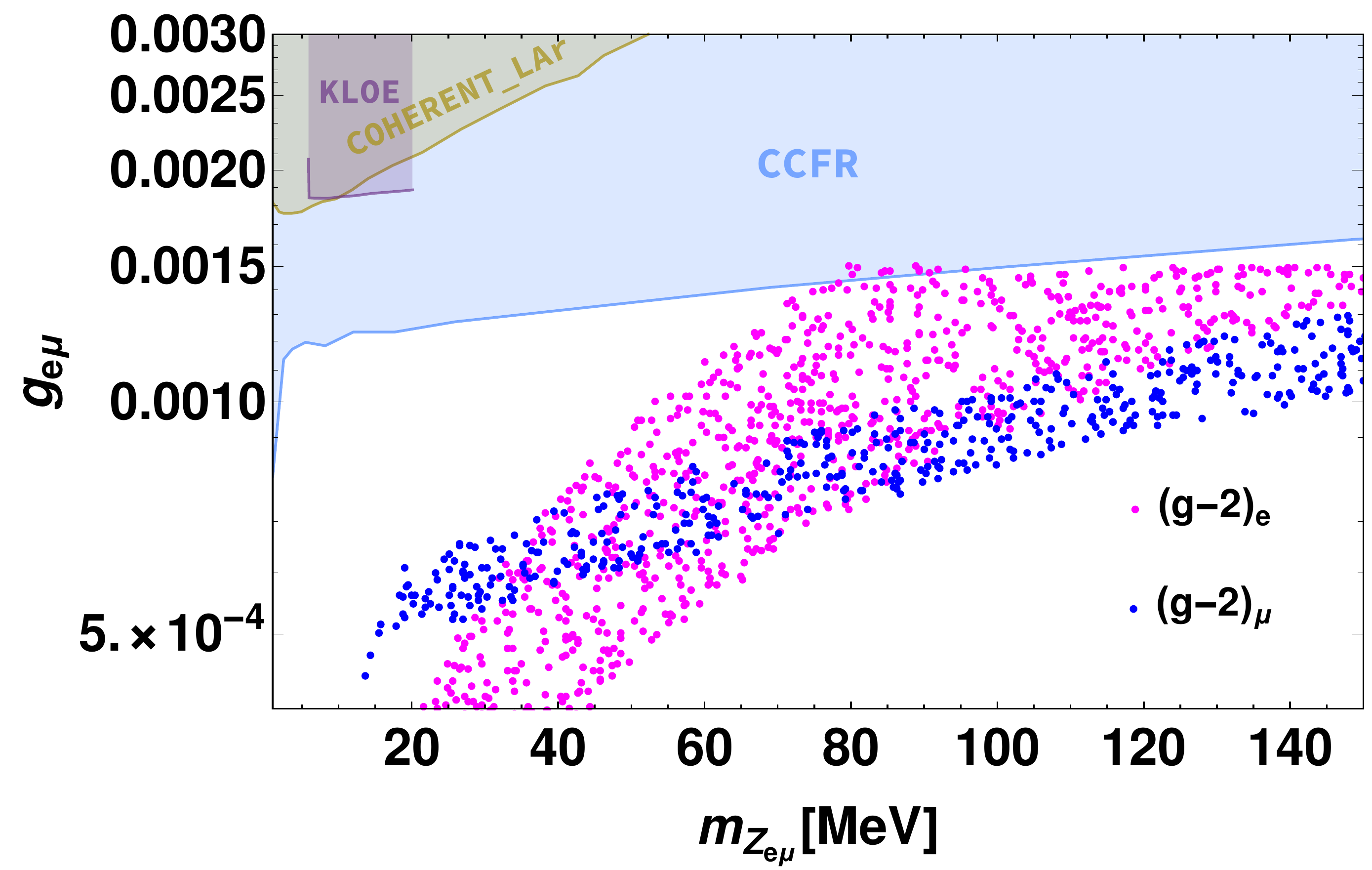}\label{electron-muon-1}}
\caption{Panel \textbf{(\ref{amug-1})} shows the variation of electron anomalous magnetic moment, $\Delta a_e$ with gauge boson mass $m_{Z_{e \mu}}$, panel \textbf{(\ref{amug-2})} shows the change of magnetic moment of muon $(\Delta a_\mu)$ with gauge boson mass $m_{Z_{e \mu}}$ (in MeV). The order of gauge coupling coming out from the figure is $~ 10^{-5}$, panel \textbf{(\ref{electron-muon-1})} shows the variation of gauge coupling $g_{e \mu}$ with gauge boson mass $m_{Z_{e \mu}}$. In the figure,  magenta scatter points satisfy electron anomalous magnetic moment while blue points give the parameter space for muon anomalous magnetic moment $(g-2)_\mu$. The limits from the experiments KLOE \cite{KLOE-2:2014qxg}, COHERENT$\_$LAr\cite{COHERENT:2017ipa} and CCFR\cite{CCFR:1991lpl} have been shown by violet, light green and light blue shaded regions respectively.}
\label{muon-1}
\end{center}
\end{figure}

The anomalous magnetic moment for electron remains an open question so far. The  value of $(g-2)_e$ is still not accurately measured, unlike $(g-2)_\mu$ with a confusion related to its sign. The determination of fine structure constant with Rubidium atom  gives us $+$ve value of $(g-2)_e$ with a discrepancy of $1.6\sigma$  over SM \cite{Morel:2020dww}, 
\begin{equation}
(\Delta a_e)_{\rm Rb} = ( 48 \pm 30) \times 10^{-14},
\label{del_aep}
\end{equation}  
whereas, for Cesium atom, \cite{Davoudiasl:2018fbb} $\Delta a_e$ has $2.4 \sigma $ discrepancy over SM as
\begin{equation}
(\Delta a_e)_{\rm Cs} = (- 87 \pm 36) \times 10^{-14} .
\label{del_aen}
\end{equation}
\begin{table}
\begin{center}
\begin{tabular}{|c|c|}
 \hline
 Parameters  &  Ranges \\
 \hline
$g_{B-L}$   &~ $[0.01,1] $~  \\ 
  $g_{e \mu}$  &  ~$[10^{-5},10^{-3}]$ ~ \\
 $v_1$ (in GeV) & ~ $[1,100] \times 10^{3}$~  \\
  $v_2$ (in GeV) & ~ $[0.3,10] \times 10^{2} $~\\
 \hline  
 \end{tabular}
 \caption{Ranges of parameters for calculating muon and electron $(g-2)$.}
 \label{muon-e-g-2} 
 \end{center}
\end{table}  
As there is an uncertainty in the extracted value of $(g-2)_e$, we will go with a more acceptable value of Rubidium atom measurement \cite{Morel:2020dww}, where $(g-2)_e$ value is +ve in sign. The Lagrangian, that provides additional  contributions to $(g-2)_e$ due to the new gauge bosons $Z_{B-L}$ and $Z_{e \mu}$ is given as 
\begin{equation}
\mathcal{L} = g_{B-L}~ \bar{e}  \gamma^\mu e~ (Z_{B-L})_\mu + g_{e \mu}~ \bar{e}  \gamma^\mu e~(Z_{e \mu})_\mu\;,
\end{equation}
with $g_{B-L} = ({\sqrt 2}~ m_{Z_{B-L}}/v_1)~ $ and $g_{e \mu} = ({\sqrt 2}~m_{Z_{e \mu}}/v_2)$ being the  gauge couplings associated with $U(1)_{B-L}$ and $U(1)_{L_e-L_\mu}$  symmetries respectively. The corresponding Feynman diagrams  are depicted in Fig. \ref{feyn} (with $l=e$), where  the left diagram represents the interaction occurring through $Z_{B-L}$ while the right one is   through $Z_{e \mu}$. The new contributions to $(g-2)_e$ arising from these diagrams are given as \cite{Moore:1984eg,Bardeen:1972vi}.
\begin{equation}
\Delta a_e = \int_0^1 \left(\frac{g^{2}_{B-L}}{4 \pi ^2} \frac{x^2 (1-x)}{x^2 + \frac{m^2_{Z_{B-L}}}{m^2_e}(1-x)} +\frac{g^{2}_{e \mu}}{4 \pi ^2} \frac{x^2 (1-x)}{x^2 + \frac{m^2_{Z_{e \mu}}}{m^2_e}(1-x)} \right) dx,
\label{dae-expre}
\end{equation}
with $x$ as the Feynman parameter. 
Using the values of gauge boson masses from Table \ref{muon-e-g-2},  it can be seen that the $(g-2)_e$ contribution coming from $Z_{B-L}$ is negligibly small,  due to the large mass of $Z_{B-L}$, i.e.,  $  \mathcal{O}(\rm TeV)$. So in eqn. \ref{dae-expre}, the main contribution towards electron $(g-2)$ comes from MeV range gauge boson, $Z_{e \mu}$.  Panel~ (\ref{amug-1}) shows the dependence of electron $(g-2)$ w.r.t. gauge boson mass $m_{Z_{e \mu}}$. It is evident that $g_{e \mu}$ should be of the order $10^{-5} $ for having the gauge boson mass $m_{Z_{e \mu}}$  in MeV range, for satisfying the electron $(g-2)$.

\subsection{Muon $(g-2)$}
Numerous studies are available in the literature illustrating muon $(g-2)$, both in the context of SM and various BSM scenarios, see refs. \cite{Aoyama:2020ynm,Bodas:2021fsy,Kamada:2018zxi,Araki:2014ona}. The recent measurement from  the MEG collaboration at Fermilab \cite{Muong-2:2021ojo} shows  a discrepancy of $4.2\sigma $ in muon anomalous magnetic moment  with respect to its SM result 
\begin{equation}
\Delta a^{\rm FNAL}_\mu = a^{\rm exp}_\mu - a^{\rm SM}_\mu =  (25.1 \pm 5.9 ) \times 10^{-10}.
\end{equation}
Previously, the result from Brookhaven National Laboratory \cite{Charity:2018sqj}, showed a discrepancy of $3.3 \sigma $  from SM 
\begin{equation}
\Delta a^{\rm BNL}_\mu =  a^{\rm exp}_\mu - a^{\rm SM}_\mu = (26.1 \pm 7.9) \times 10^{-10}.
\end{equation}
Hence, to accommodate the above  deviation, one needs to go beyond SM. In the context of the present model,   the new  interactions of muon involving $Z_{B-L}$ and $ Z_{e \mu} $ bring the extra contributions towards $(g-2)_\mu$. 
The relevant Feynman diagrams are shown in Fig. \ref{feyn} with ($l=\mu$) and the corresponding neutral current interaction Lagrangian is given as
\begin{eqnarray}
\mathcal{L} =  g_{B-L} ~\bar{\mu} \gamma^\mu \mu~({Z_{B-L}})_\mu - g_{e \mu} ~\bar{\mu}  \gamma^\mu \mu~({Z_{e \mu}})_\mu ,
\end{eqnarray}
with $g_{B-L} $ and $g_{e \mu}$ are the two associated gauge couplings respectively. Similar to $(g-2)_e$, the anomalous magnetic moment of muon is also dominantly depends on $Z_{e \mu}$ gauge boson.
The expression for $\Delta a_\mu$ is given as follows \cite{Moore:1984eg,Bardeen:1972vi},
\begin{equation}
\Delta a_\mu = \int_0^1 \left(\frac{g^{2}_{B-L}}{4 \pi ^2} \frac{x^2 (1-x)}{x^2 + \frac{m^2_{Z_{B-L}}}{m^2_\mu}(1-x)} +\frac{g^{2}_{e \mu}}{4 \pi ^2} \frac{x^2 (1-x)}{x^2 + \frac{m^2_{Z_{e \mu}}}{m^2_\mu}(1-x)} \right) dx.
\label{damu-exp}
\end{equation}
 In eqn. \ref{damu-exp}, we can safely ignore the  first term involving  the contribution from $U(1)_{B-L}$ gauge boson,  as the mass of corresponding gauge boson  is  quite heavy. Panel (\ref{amug-2}) represents the dependence of $\Delta a_{\mu}$ with the gauge boson mass $m_{Z_{e \mu}}$. 
 
  Finally, we have  shown the allowed parameter space in $g_{e \mu}-m_{Z_{e\mu}}$ plane, for both electron and muon anomalous magnetic moments simultaneously. Panel~ (\ref{electron-muon-1}) shows the preferred region for $g_{e \mu}-m_{Z_{e \mu}}$ for both $\Delta a_e$ and $\Delta a_\mu$. In this panel,  the  points represented by magenta color satisfy electron anomalous magnetic moment while points in blue  give the parameter space for muon anomalous magnetic moment. The limits from the experiments, KLOE \cite{KLOE-2:2014qxg}, COHERENT$\_$LAr\cite{COHERENT:2017ipa} and CCFR\cite{CCFR:1991lpl} are depicted within the panel. From the panel (\ref{electron-muon-1}), we can observe that our model successfully explains both muon and electon $(g-2)$ simultaneously and the allowed $g_{e \mu}-m_{Z_{e\mu}}$ parameter space is   consistent with the constraints obtained from the experiments: KLOE, COHERENT$\_$LAr and CCFR.

\vspace*{0.25 true cm}


\section{Leptogenesis}
\label{sec:lepto}
\begin{figure}[htbp]
\begin{center}
\includegraphics[height=65mm, width=85mm]{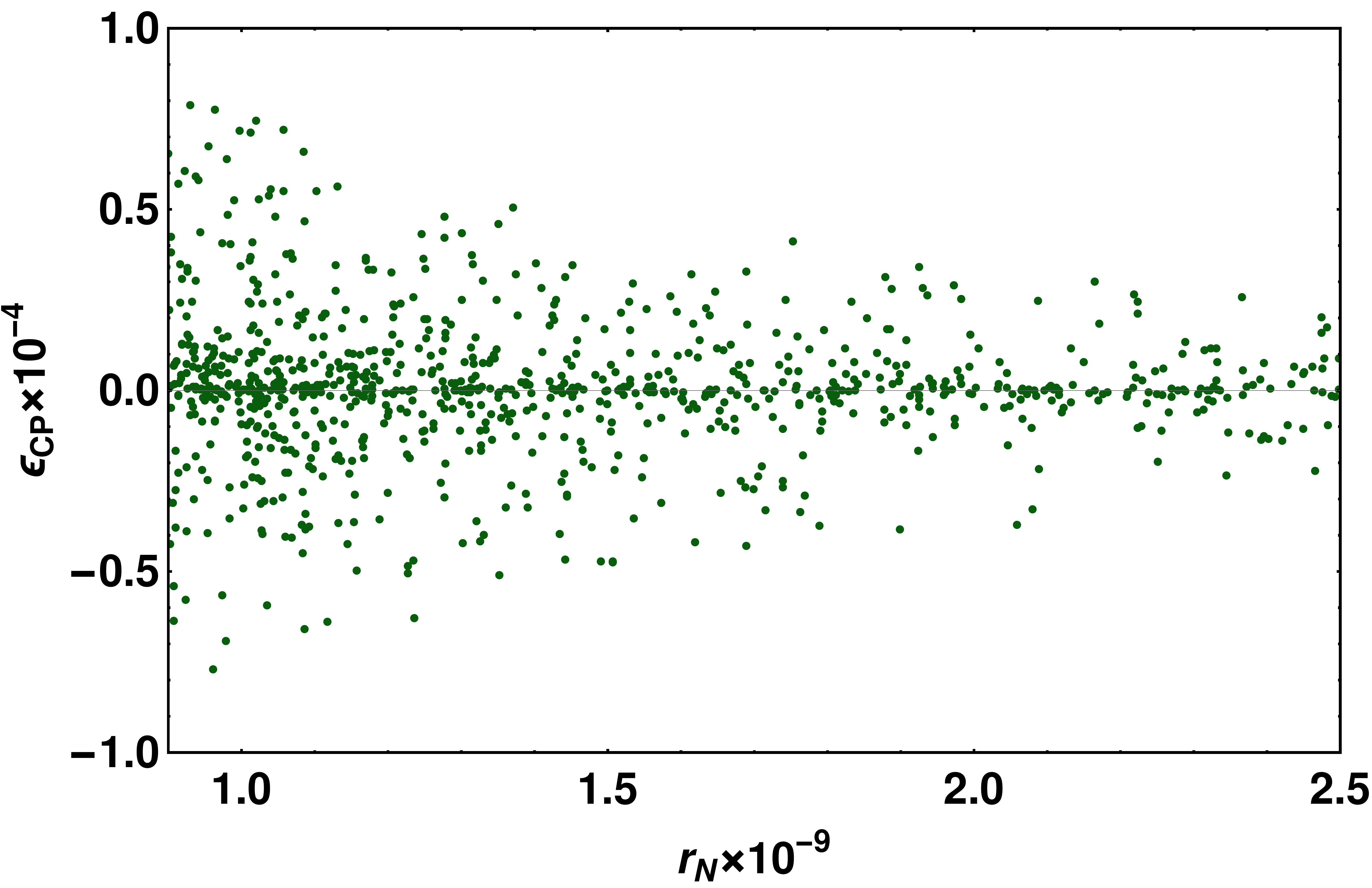}
\caption{Variation of CP asymmetry with respect to $r_N$ parameter.
}
\label{cp-asym}
\end{center}
\end{figure}

Generating the observed baryon asymmetry of the Universe can be accomplished through leptogenesis, which is considered to be one of the most favored methods. By utilizing the standard scenario of resonant enhancement in CP asymmetry, the scale has been reduced to as low as TeV \cite{Pilaftsis:1997jf,Bambhaniya:2016rbb,Pilaftsis:2003gt, Abada:2018oly,Buchmuller:2004nz}. The present model is the case of inverse seesaw where in the basis of $(N^c_{R_i}, S_{L_i})$, we have the heavy fermion mass matrix of the form given in eqn. (\ref{lepto-mass-1}). The presence of $\mathcal{M}_\mu$ term  acts like a small mass splitting term, 
\begin{eqnarray}
\quad \mathcal{M} = \begin{pmatrix}
 0  &  \mathcal{M_{NS}}  \cr \mathcal{M_{NS}^T}  &  \mathcal{M_{\mu}} \end{pmatrix}.
 \label{lepto-mass-1}
\end{eqnarray} 
The typical size of  $\mathcal{M}_\mu \ll \mathcal{M_{NS}}$, and block diagonalization of $\mathcal{M}$ as in eqn.~\ref{lepto-mass-1} by using unitary matrix $\frac{1}{\sqrt{2}} \begin{pmatrix}
 \textit{I}  &  -\textit{I}  \cr \textit{I}  &  \textit{I} \end{pmatrix}$, yields a diagonal $\mathcal{M}^\prime$ matrix as shown in eqn.~\ref{lepto-mass}.
 \begin{eqnarray}
\quad \mathcal{M^{\prime}} = \begin{pmatrix}
 \mathcal{M_{NS}} + \frac{\mathcal{M_{\mu}}}{2}  &  -\frac{\mathcal{M_{\mu}}}{2}  \cr -\frac{\mathcal{M_{\mu}}}{2}     &  -\mathcal{M_{NS}}+\frac{\mathcal{M_{\mu}}}{2} \end{pmatrix}  & 
\approx  &  \begin{pmatrix}
 \mathcal{M_{NS}} + \frac{\mathcal{M_{\mu}}}{2}  &  0  \cr 0     &  -\mathcal{M_{NS}}+\frac{\mathcal{M_{\mu}}}{2} \end{pmatrix}. 
 \label{lepto-mass}
\end{eqnarray} 
Now, the mass eigenstates can be related to flavour eigenstates through the unitary rotation matrix as
\begin{eqnarray}
\begin{pmatrix}
    S_{L_i} \cr
    N_{R_i}^c
\end{pmatrix} = \begin{pmatrix}
 \cos \theta  &  -\sin \theta  \cr \sin \theta  &  \cos \theta \end{pmatrix} \begin{pmatrix}
     N_{i}^+   \cr N_i^- 
 \end{pmatrix}. 
 \label{eigen}
\end{eqnarray} 
From eqn. \ref{eigen}, in the limit of maximal mixing, the additional fermions can be related to their mass eigenstates by the expressions \cite{Aoki:2015owa}
\begin{eqnarray}
    N_{R_i}^c = \frac{(N_i^+ + N_i^-)}{\sqrt{2}}\;, ~~~~~~~
    S_{L_i} = \frac{(N_i^+ - N_i^-)}{\sqrt{2}}\;.
\end{eqnarray}
Thus, one can write the  Lagrangian in mass basis of the heavy fermions as 
\begin{eqnarray}
    \mathcal{L_{\rm lepton}} \supset Y_S \bar{l}_{L} \tilde{H} N_{i}^+ + Y_N \bar{l}_{L} \tilde{H} N_{i}^- + \frac{\mathcal{M_{NS}}-\frac{\mathcal{M_{\mu}}}{2}}{2} \bar{N}_i^{- c} N_i^- + \frac{\mathcal{M_{NS}}+\frac{\mathcal{M_{\mu}}}{2}}{2} \bar{N}_i^{+ c} N_i^+ + {\rm h.c.}
\end{eqnarray}
The mass eigenvalues for new states $N_i^+, N_i^-$ can be obtained by diagonalizing the matrix \ref{diag_lepto_mass}, 
\begin{eqnarray}
    \mathcal{M_{NS}} \pm 
    \frac{\mathcal{M_{\mu}}}{2} = \begin{pmatrix}
     \frac{v_1}{\sqrt{2}} |y_N^1| e^{ i \phi_2} &  ~~  \pm \frac{\mathcal{M}_{12}}{2} & ~~  \pm y_{13} \frac{v_2}{2 \sqrt{2}} \cr \pm \frac{\mathcal{M}_{12}}{2} & ~~  \frac{v_1}{\sqrt{2}} y_N^{2} & ~~ \pm y_{23} \frac{v_2}{2 \sqrt{2}} \cr \pm y_{13} \frac{v_2}{ 2 \sqrt{2}} & ~~ \pm y_{23} \frac{v_2}{\sqrt{2}} & ~~ \frac{v_1}{\sqrt 2}y_N^3 \pm \frac{\mathcal{M}_{33}}{2}
    \end{pmatrix}\;. \label{diag_lepto_mass}
\end{eqnarray}
The numerical computation of the six mass eigenvalues results in the identification of $N_1^-$ as the particle with the lowest mass, as presented in Table \ref{lepto-benchmark}. Additionally, we assume that the contribution to the CP asymmetry is mainly due to the lightest pair of particles with masses at the TeV scale. The negligible difference between the masses of these lightest particles implies that the contribution from the self-energy of heavy particle decay through  one-loop is more significant than the contribution coming from the vertex diagram. The formula for calculating the CP asymmetry is expressed in the following manner according to \cite{Gu:2010xc},
\begin{eqnarray}
({\epsilon_{\rm CP}})_{\pm} &=& \frac{\Gamma (N_1^{\pm} \rightarrow \ell H) - 
    \Gamma ( N_1^{\pm} \rightarrow \bar{\ell} H^{\dagger} )}{\Gamma (N_1^{\pm} \rightarrow \ell H) + 
    \Gamma ( N_1^{\pm} \rightarrow \bar{\ell} H^{\dagger} )} 
    \approx \frac{\mathrm{Im} ( Y_N^{\dagger} Y_S Y_{N}^{\dagger} Y_S)_{11}}{8 \pi A_{\pm}} \frac{r_N}{r_N^2 + \Gamma^2_{\pm} / m_{N_1^{\pm}}^2}\;,
\end{eqnarray}
where, $r_N=\displaystyle{\frac{m_{N_1^{+}}^2 ~-~ m_{N_1^{-}}^2}{m_{N_1^+}~m_{N_1^-}}}$ = $\displaystyle{\frac{ \Delta m \left(m_{N_1^+}~+~m_{N_1^-}\right)}{m_{N_1^+} ~m_{N_1^-}}}$ with $A_{+} = (Y_S^{\dagger} Y_S)_{11}$, $A_-=(Y_N^{\dagger} Y_N)_{11}$. 
$\Gamma_{\pm} = \displaystyle{A_{\pm} \frac{m_{N_1^{\pm}}}{8 \pi}}$ being the decay width of $N_1^\pm$ and note that at the leading order, the matrix elements of $Y_N$ and $Y_S$ are approximately of the same magnitude as $y_D^{\alpha}$ ( $\alpha=e,\mu,\tau$), i.e., $Y_N \approx Y_S \approx \displaystyle{\frac{1}{\sqrt{2}}}~y_D^{\alpha}$.  Therefore, the aforementioned details enable us to illustrate the CP asymmetry plots, as depicted in Fig.~\ref{cp-asym}, which displays  the correlation of the $r_N$  parameter with $\epsilon_{\rm CP}$.
Subsequently, we will discuss the benchmark values, presented in Table \ref{lepto-benchmark}, which satisfy both the neutrino mass and the requisite CP asymmetry for getting the  lepton asymmetry that can generate the observed baryon asymmetry.

The dynamics of the relevant Boltzmann equations can be used to deduce the evolution of lepton asymmetry. According to the Sakharov criteria \cite{Sakharov:1967dj}, the decay of the parent fermion must be out of equilibrium to produce the lepton asymmetry. In order to satisfy this condition, the Hubble rate must be compared with the decay rate, as shown below
\begin{equation}
    K_{N_1^-} = \frac{\Gamma_{N_1^-}}{H (T = m_{N_1^-})}\;,
\end{equation}
where $H = \displaystyle{\frac{1.67 \sqrt{g_\star}~ T^2 }{M_{\rm Pl}}}$, with $g_{\star}= 106.75 $, and $M_{\rm Pl}$ is $1.22 \times 10^{19}$ GeV. We further assume that the coupling strength is approximately equal to $\left(\approx y_D^e \right)$ and of the order of $10^{-7}$, which is consistent with the minimum order of coupling parameters obtained from the numerical analysis section and in accordance with neutrino oscillation data. This results in $K_{N_1^-} \sim 1$, preventing the inverse decay from reaching thermal equilibrium. The Boltzmann equations, which detail the evolution of the right-handed fermion and lepton number densities represented in terms of the yield parameter (the ratio of the number density to entropy density), are provided in \cite{Plumacher:1996kc, Giudice:2003jh, Strumia:2006qk, Iso:2010mv}. 
\begin{eqnarray}
&& \frac{d Y_{N_1^-}}{dz}=-\frac{z}{s H(m_{N_1^-})} \left[\left( \frac{Y_{N_1^-}}{{Y^{\rm eq}_{N_1^-}}}-1\right)\gamma_D +\left( \left(\frac{{Y_{N_1^-}}}{{Y^{\rm eq}_{N_1^-}}}\right)^2-1\right)\gamma_S \right],\label{Boltz1}\\
&& \frac{d Y_{B-L}}{d z}= -\frac{z}{s H(m_{N_1^-})} \gamma_D \left[ \epsilon_{N_1^-} \left( \frac{Y_{N_1^-}}{{Y^{\rm eq}_{N_1^-}}}-1\right)  + \frac{Y_{B-L}}{{2Y^{\rm eq}_{\ell}}} \right].
\label{Boltz}
\end{eqnarray}
Here, $s$ denotes the entropy density, $z = m_{N_1^-}/T$, and the equilibrium number densities are outlined in \cite{Davidson:2008bu} as
\begin{eqnarray}
Y^{\rm eq}_{N_1^-}= \frac{45  g_{N_1^-}}{4 {\pi}^4 g_\star} z^2 K_2(z), \hspace{3mm} {Y^{\rm eq}_\ell}= \frac{3}{4} \frac{45 \zeta(3) g_\ell}{2 {\pi}^4 g_{\star}}\,.
\end{eqnarray}
Here,   $g_{N_1^-}=2$ and $g_\ell=2$ denote the degrees of freedom of right-handed fermions and leptons, respectively. The decay rate is given by $\gamma_D = s Y^{\rm eq}_{N_1^-}\Gamma_D$ where, $\Gamma_{D} = \Gamma_{N_1^-} \frac{K_1(z)}{K_2(z)}$ with $K_{1,2}$ denote modified Bessel functions. While $\gamma_S$ denotes the scattering rate of the decaying fermion, and  the Boltzmann eqn. for $Y_{B-L}$ is    free    from    the    subtlety    of    asymmetry    getting produced even when $N_1^-$ is in thermal equilibrium, i.e., by subtracting the on-shell $N_1^-$ exchange contribution ($\frac{\gamma_{D}}{4}$) from the $\Delta L =2$ processes \cite{Giudice:2003jh}.
 \begin{figure}[htbp]
\begin{center}
\subfloat[]{\includegraphics[height=65mm, width=75mm]{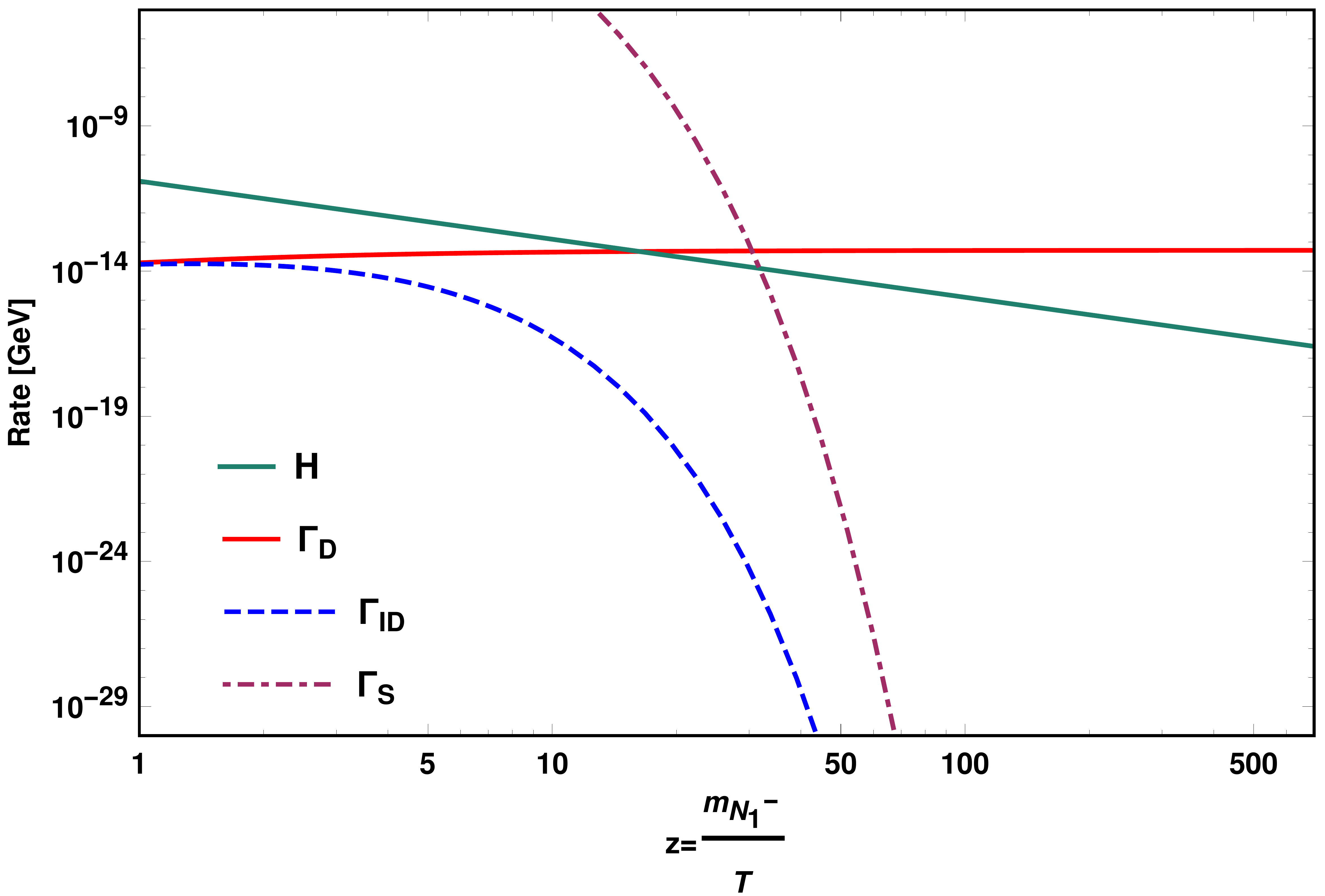}\label{rate_lepto}}
\hspace*{0.3 true cm}
\subfloat[]{\includegraphics[height=65mm, width=80mm]{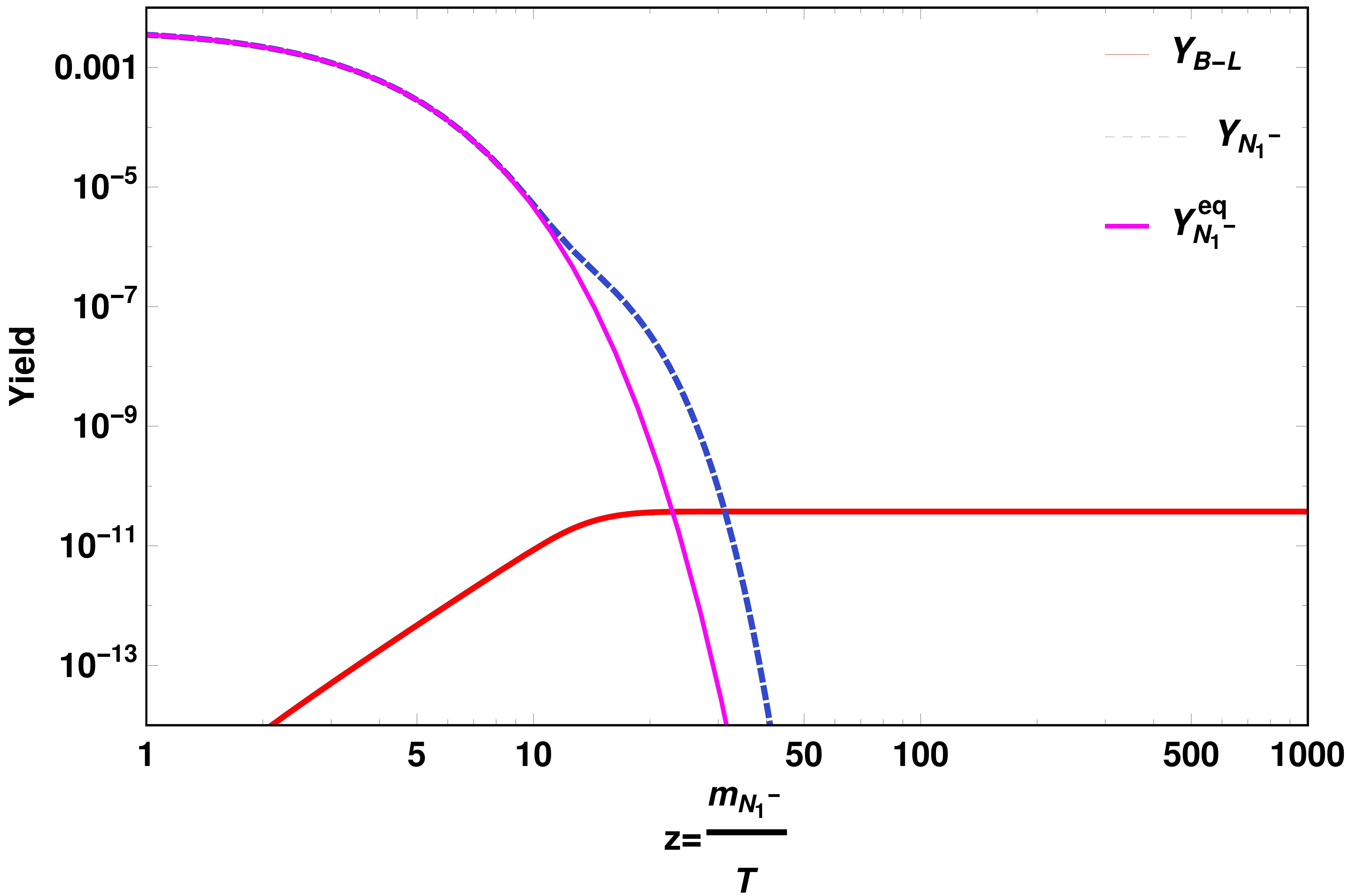}\label{yield-lepto}}
\caption{ Plot (\textbf{\ref{rate_lepto}}) projects the   comparison of interaction rates with Hubble expansion (green line), where the solid red line corresponds to decay, inverse decay (blue dashed), and scattering rate shown by brown dot-dashed line, whereas, plot (\textbf{\ref{yield-lepto}}) exhibits the evolution of $Y_{B-L}$(red solid) as a function of $z=m_{N_1^-}/T$.}
\label{boltz-asym}
\end{center}
\end{figure}
In the panel (\ref{rate_lepto}), the interaction rates are compared to the Hubble expansion (green solid line), the decay rate ($\Gamma_D$, red solid line), and the inverse decay rate $\left( \Gamma_{ID} = \Gamma_{D}\frac{Y^{eq}_{N_1^-}}{Y^{eq}_\ell}\right)$, depicted by a dashed blue line. It is observed that the inverse decay does not reach thermal equilibrium. The scattering rate for decaying fermion $\left(\frac{\gamma_S}{s Y^{\rm eq}_{N_1^-}}\right)$ is projected by the brown dot-dashed line and is consistent with neutrino oscillation studies. Once the out-of-equilibrium criteria are met, the decay proceeds slowly, resulting in over-abundance of $Y_{N_1^-}$ (blue dashed curve in panel \ref{yield-lepto}), which does not follow $Y^{\rm eq}_{N_1^-}$ (magenta solid curve), and lepton asymmetry (red solid curve) is generated. The obtained lepton asymmetry is then converted into the observed baryon asymmetry through the sphaleron transition, as outlined in \cite{Plumacher:1996kc, Planck:2015fie,Cyburt:2015mya}.
\begin{equation}
Y_B = \left(\frac{8N_f + 4 N_H}{22 N_f + 13 N_H}\right)Y_{B-L}.
\end{equation}
Here, $N_f$ denotes the number of fermion generations, and $N_H$ is the number of Higgs doublets.
The observed baryon asymmetry is quantified in terms of baryon to photon ratio \cite{Planck:2018vyg}
\begin{equation}
\eta = \frac{\eta_b - \eta_{\bar{b}}}{\eta_\gamma} = 6.08 \times 10^{-10}.
\end{equation}
Based on the relation $Y_{B} = (7.04)^{-1} \eta$, the current bound on baryon asymmetry is $Y_{B} \sim 8.6\times 10^{-11}$. Taking the asymptotic value ($10^{-11}$) of $Y_{B-L}$ from the panel (\ref{yield-lepto}) (red solid), the obtained baryon asymmetry is $Y_{B} = \frac{28}{79}~ Y_{B-L} \sim 10^{-11}$.

\begin{table}[]
    \centering
    \begin{tabular}{|c|c|c|c|}
    \hline
        Parameters & $M_{N_1^-}$ (GeV) &  $\epsilon_{CP}$ & $\Delta m (\rm{GeV}) $ \\
        \hline
         BP & $2.96 \times 10^{3}$  &    $1.5 \times 10^{-5}$  &  $1.5 \times 10^{-6}$ \\
         \hline
    \end{tabular}
    \caption{In this table, we show the benchmark points (BP) of the mutual parameter space satisfying neutrino oscillation data for giving the correct lepton asymmetry.}
    \label{lepto-benchmark}
\end{table}


\section{comments on Collider bound}
\label{coll-bound}

\begin{figure}[htbp]
    \centering
    \includegraphics[height=60mm, width=88mm]{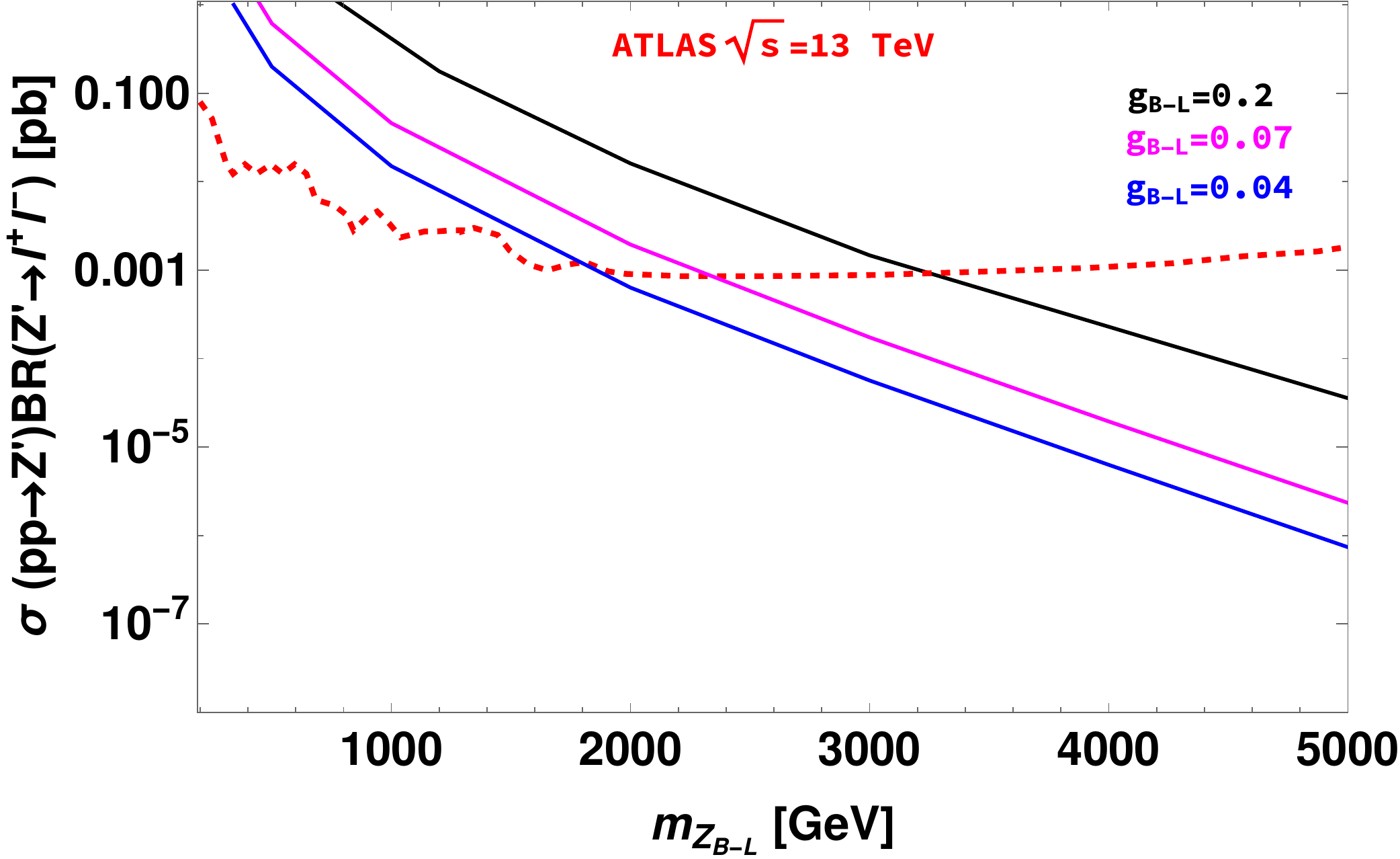}
    \caption{Collider search by ATLAS dilepton constraints on the proposed model is shown. Here,
the red dashed line represents the exclusion limit from ATLAS, and blue, magenta, and black curves show the dilepton signal cross sections for gauge coupling values as 0.04, 0.07, and 0.2, respectively, with the function of $m_{Z_{B-L}}$.}
    \label{coll-bound-fig}
\end{figure}

In current and past collider physics programs, so many experiments have given the upper limits on the mass of the new gauge bosons with respect to the corresponding gauge couplings. Most of these experiments have searched for new heavy resonances in both dilepton and dijet signals. Among these experiments, ATLAS put a stronger bound on the dilepton signal than dijet due to small background events. In Fig.~\ref{coll-bound-fig}, the upper bound of ATLAS limit for the cross-section of $pp \rightarrow Z_{B-L} \rightarrow  \ell \ell$ $(\ell = e, \mu)$ with respect to heavy gauge boson mass has been shown by a red dotted curve with the center of mass frame momentum $\sqrt{s}= 13$ TeV. In our proposed model, gauge boson associated with $U(1)_{B-L}$ gauge symmetry can be tested by current and future collider experiments. From Fig.~\ref{coll-bound-fig}, it can be seen that for $g_{B-L}=0.2$ (black curve), the range $m_{Z_{B-L}} \geq 3.4 ~ \rm TeV$ is excluded, while for $g_{B-L} = 0.07$ (magenta curve) and $g_{B-L} = 0.04$ (blue curve) the mass range $m_{Z_{B-L}} \geq 2.5 ~ \rm{TeV}$ and $m_{Z_{B-L}} \geq 1.5 ~ \rm {TeV}$ are excluded. Thus, as we can go greater values of $g_{B-L}$, the corresponding allowed parameter space increases. 
\section{Summary and Conclusion}
\label{sec: concl}
This article comprehensively analyzes various aspects of neutrino phenomenology, along with electron and muon $(g-2)$, resonant leptogenesis, and probing the model in future long-baseline experiments within an inverse seesaw framework. The model includes three additional right-handed neutrinos ($N_{R_i}$) and three neutral fermions ($S_{L_i}), ~i=1,2,3$, with the required quantum numbers to cancel the gauge anomalies via two local gauge symmetries $U(1)_{B-L}$ and $U(1)_{L_e-L_\mu}$. The spontaneous breaking of these symmetries through the VEVs of the scalars $\chi_1$ and $\chi_2$ induce masses  to the scalar sector as well as gauge bosons. The study shows that the sum of active neutrino masses ($\sum m_{i}$) is consistent with normal ordering, varies from 0.058 eV to 0.061 eV. Additionally, the oscillation parameters such as $\sin^2 \theta_{23}$ strongly prefer the upper octant and vary only within the range of $[0.5,0.6]$, while $\sin^2 \theta_{12}$ ranges from $[0.27,0.34]$, and $\sin^2 \theta_{13}$ varies from $[0.02029,0.02389]$. The differences in mass squared values are also altered to $\Delta m_{21}^2 = [6.82,8.03] \times 10^{-5}$ eV$^2$ and $\Delta m_{31}^2 =[2.43,2.59]\times 10^{-3}$  eV$^2$. The CP-violating phase has a more restricted range, with a value of $[180^{\circ},224^{\circ}]$ based on our model. Table \ref{best-fit-nufit} lists the best-fit values for oscillation parameters in our model. Furthermore, we examine our model in future long-baseline experiments such as DUNE, T2HK, and T2HKK. Results from this subsection conclude that $3\sigma$ allowed region  of $\theta_{23}-\delta_{\rm CP}$ plane can be tested by $3\sigma$ as well as $5\sigma$ allowed range of DUNE. 
Similarly, the allowed region of $3\sigma$ C.L. is consistent with $5\sigma$ C.L. of T2HK. 
But, a small portion of $2\sigma$ exclusion region is not compatible with $3\sigma$ of T2HK. $5\sigma$ region of T2HKK can test our model with $3\sigma$ allowed range of our proposed model. To account for the anomalous magnetic moments of electrons and muons, we include the contributions from two gauge bosons ($Z_{B-L}$ and $Z_{e \mu}$) in the expressions of $\Delta a_e$ and $\Delta a_\mu$. We found that only $Z_{e \mu}$ is involved in the calculation of anomalous magnetic moment, as $m_{Z_{B-L}}$ is in the TeV range. Furthermore, our model is capable of simultaneously explaining the anomalous magnetic moments of electrons and muons, satisfying the current results. The heavy fermions flavor structure  would have resulted in six doubly degenerate mass eigenstates. However, the presence of $\mathcal{M}_\mu$,  as discussed above in the neutrino phenomenology section, saves the day by acting like a small mass-splitting term, making it possible to explain leptogenesis. The lightest heavy fermion eigenstate's decay produced a non-zero CP asymmetry. The self-energy contribution is partially enhanced due to the slight mass difference between the two lighter heavy fermions. We used a specific set of model parameters consistent with oscillation data to solve coupled Boltzmann equations. This approach enabled us to obtain the evolution of lepton asymmetry at the TeV scale, which is of the order of  $\approx 10^{-11}$. This value is sufficient to explain the current baryon asymmetry of the Universe. Finally, the collider bound places a stringent constraint on the parameter space of dilepton production cross section and $m_{Z_{B-L}}$.


\section*{Acknowledgments}
PP and PM want to thank Prime Minister's Research Fellowship (PMRF) scheme for its financial support. MKB wants to thank DST-Inspire for financial support. RM would like to acknowledge University of Hyderabad IoE project grant no. RC1-20-012. We gratefully acknowledge the use of CMSD HPC facility of Univ. of Hyderabad to carry out the computational work.


\bibliographystyle{JHEP}
\bibliography{gauge}

\providecommand{\href}[2]{#2}\begingroup\raggedright\begin{thebibliography}{10}

\bibitem{Ma:1998dn}
E.~Ma, {\it {Pathways to naturally small neutrino masses}},  {\em Phys. Rev.
  Lett.} {\bf 81} (1998) 1171--1174,
  [\href{http://arxiv.org/abs/hep-ph/9805219}{{\tt hep-ph/9805219}}].

\bibitem{Weinberg:1979sa}
S.~Weinberg, {\it {Baryon and Lepton Nonconserving Processes}},  {\em Phys.
  Rev. Lett.} {\bf 43} (1979) 1566--1570.

\bibitem{Ma:2005py}
E.~Ma, {\it {Connection between the neutrino seesaw mechanism and properties of
  the Majorana neutrino mass matrix}},  {\em Phys. Rev. D} {\bf 71} (2005)
  111301, [\href{http://arxiv.org/abs/hep-ph/0501056}{{\tt hep-ph/0501056}}].

\bibitem{Palcu:2009uk}
A.~Palcu, {\it {Canonical Seesaw Mechanism in Electro-Weak SU(4)(L) x U(1)(Y)
  Models}},  {\em Mod. Phys. Lett. A} {\bf 24} (2009) 2589--2600,
  [\href{http://arxiv.org/abs/0908.1636}{{\tt arXiv:0908.1636}}].

\bibitem{Zhang:2011vh}
H.~Zhang, {\it {Light Sterile Neutrino in the Minimal Extended Seesaw}},  {\em
  Phys. Lett. B} {\bf 714} (2012) 262--266,
  [\href{http://arxiv.org/abs/1110.6838}{{\tt arXiv:1110.6838}}].

\bibitem{DiValentino:2019dzu}
E.~Di~Valentino, A.~Melchiorri, and J.~Silk, {\it {Cosmological constraints in
  extended parameter space from the Planck 2018 Legacy release}},  {\em JCAP}
  {\bf 01} (2020) 013, [\href{http://arxiv.org/abs/1908.01391}{{\tt
  arXiv:1908.01391}}].

\bibitem{Rodejohann:2004cg}
W.~Rodejohann, {\it {Type II seesaw mechanism, deviations from bimaximal
  neutrino mixing and leptogenesis}},  {\em Phys. Rev. D} {\bf 70} (2004)
  073010, [\href{http://arxiv.org/abs/hep-ph/0403236}{{\tt hep-ph/0403236}}].

\bibitem{Ding:2017jdr}
R.~Ding, Z.-L. Han, L.~Feng, and B.~Zhu, {\it {Confronting the DAMPE Excess
  with the Scotogenic Type-II Seesaw Model}},  {\em Chin. Phys. C} {\bf 42}
  (2018), no.~8 083104, [\href{http://arxiv.org/abs/1712.02021}{{\tt
  arXiv:1712.02021}}].

\bibitem{deSousaPires:2018fnl}
C.~A. de~Sousa~Pires, F.~Ferreira De~Freitas, J.~Shu, L.~Huang, and P.~Wagner
  Vasconcelos~Oleg\'ario, {\it {Implementing the inverse type-II seesaw
  mechanism into the 3-3-1 model}},  {\em Phys. Lett. B} {\bf 797} (2019)
  134827, [\href{http://arxiv.org/abs/1812.10570}{{\tt arXiv:1812.10570}}].

\bibitem{Dong:2010gk}
P.~V. Dong, L.~T. Hue, H.~N. Long, and D.~V. Soa, {\it {The 3-3-1 model with
  $\mathrm{A}_4$ flavor symmetry}},  {\em Phys. Rev. D} {\bf 81} (2010) 053004,
  [\href{http://arxiv.org/abs/1001.4625}{{\tt arXiv:1001.4625}}].

\bibitem{Franceschini:2008pz}
R.~Franceschini, T.~Hambye, and A.~Strumia, {\it {Type-III see-saw at LHC}},
  {\em Phys. Rev. D} {\bf 78} (2008) 033002,
  [\href{http://arxiv.org/abs/0805.1613}{{\tt arXiv:0805.1613}}].

\bibitem{Goswami:2018jar}
S.~Goswami, K.~N. Vishnudath, and N.~Khan, {\it {Constraining the minimal
  type-III seesaw model with naturalness, lepton flavor violation, and
  electroweak vacuum stability}},  {\em Phys. Rev. D} {\bf 99} (2019), no.~7
  075012, [\href{http://arxiv.org/abs/1810.11687}{{\tt arXiv:1810.11687}}].

\bibitem{Biswas:2019ygr}
A.~Biswas, D.~Borah, and D.~Nanda, {\it {Type III seesaw for neutrino masses in
  U(1)$_{B-L}$ model with multi-component dark matter}},  {\em JHEP} {\bf 12}
  (2019) 109, [\href{http://arxiv.org/abs/1908.04308}{{\tt arXiv:1908.04308}}].

\bibitem{FileviezPerez:2008sr}
P.~Fileviez~Perez, {\it {Type III Seesaw and Left-Right Symmetry}},  {\em JHEP}
  {\bf 03} (2009) 142, [\href{http://arxiv.org/abs/0809.1202}{{\tt
  arXiv:0809.1202}}].

\bibitem{Mishra:2023cjc}
P.~Mishra, M.~K. Behera, and R.~Mohanta, {\it {Neutrino phenomenology, W mass
  anomaly \& muon $(g-2)$ in minimal type-III seesaw using $T^\prime$ modular
  symmetry}},  \href{http://arxiv.org/abs/2302.00494}{{\tt arXiv:2302.00494}}.

\bibitem{Mishra:2022egy}
P.~Mishra, M.~K. Behera, P.~Panda, and R.~Mohanta, {\it {Type III seesaw under
  $A_4$ modular symmetry with leptogenesis}},  {\em Eur. Phys. J. C} {\bf 82}
  (2022), no.~12 1115, [\href{http://arxiv.org/abs/2204.08338}{{\tt
  arXiv:2204.08338}}].

\bibitem{Arbelaez:2021chf}
C.~Arbel\'aez, C.~Dib, K.~Mons\'alvez-Pozo, and I.~Schmidt, {\it {Quasi-Dirac
  neutrinos in the linear seesaw model}},  {\em JHEP} {\bf 07} (2021) 154,
  [\href{http://arxiv.org/abs/2104.08023}{{\tt arXiv:2104.08023}}].

\bibitem{Behera:2020sfe}
M.~K. Behera, S.~Mishra, S.~Singirala, and R.~Mohanta, {\it {Implications of
  $A_4$ modular symmetry on Neutrino mass, Mixing and Leptogenesis with Linear
  Seesaw}},  \href{http://arxiv.org/abs/2007.00545}{{\tt arXiv:2007.00545}}.

\bibitem{Behera:2022wco}
M.~K. Behera and R.~Mohanta, {\it {Linear seesaw in $A^\prime_5$ modular
  symmetry with Leptogenesis}},  \href{http://arxiv.org/abs/2201.10429}{{\tt
  arXiv:2201.10429}}.

\bibitem{Hirsch:2009mx}
M.~Hirsch, S.~Morisi, and J.~W.~F. Valle, {\it {A4-based tri-bimaximal mixing
  within inverse and linear seesaw schemes}},  {\em Phys. Lett. B} {\bf 679}
  (2009) 454--459, [\href{http://arxiv.org/abs/0905.3056}{{\tt
  arXiv:0905.3056}}].

\bibitem{Deppisch:2015cua}
F.~F. Deppisch, L.~Graf, S.~Kulkarni, S.~Patra, W.~Rodejohann, N.~Sahu, and
  U.~Sarkar, {\it {Reconciling the 2 TeV excesses at the LHC in a linear seesaw
  left-right model}},  {\em Phys. Rev. D} {\bf 93} (2016), no.~1 013011,
  [\href{http://arxiv.org/abs/1508.05940}{{\tt arXiv:1508.05940}}].

\bibitem{Ma:2009gu}
E.~Ma, {\it {Radiative inverse seesaw mechanism for nonzero neutrino mass}},
  {\em Phys. Rev. D} {\bf 80} (2009) 013013,
  [\href{http://arxiv.org/abs/0904.4450}{{\tt arXiv:0904.4450}}].

\bibitem{Behera:2020lpd}
M.~K. Behera, S.~Singirala, S.~Mishra, and R.~Mohanta, {\it {A modular A $_{4}$
  symmetric scotogenic model for neutrino mass and dark matter}},  {\em J.
  Phys. G} {\bf 49} (2022), no.~3 035002,
  [\href{http://arxiv.org/abs/2009.01806}{{\tt arXiv:2009.01806}}].

\bibitem{Behera:2021eut}
M.~K. Behera and R.~Mohanta, {\it {Inverse seesaw in $A_5^\prime$ modular
  symmetry}},  {\em J. Phys. G} {\bf 49} (2022), no.~4 045001,
  [\href{http://arxiv.org/abs/2108.01059}{{\tt arXiv:2108.01059}}].

\bibitem{Parida:2010wq}
M.~K. Parida and A.~Raychaudhuri, {\it {Inverse see-saw, leptogenesis,
  observable proton decay and $\Delta^{\pm\pm}_{\rm R}$ in SUSY SO(10) with
  heavy W\_R}},  {\em Phys. Rev. D} {\bf 82} (2010) 093017,
  [\href{http://arxiv.org/abs/1007.5085}{{\tt arXiv:1007.5085}}].

\bibitem{Dias:2012xp}
A.~G. Dias, C.~A. de~S.~Pires, P.~S. Rodrigues~da Silva, and A.~Sampieri, {\it
  {A Simple Realization of the Inverse Seesaw Mechanism}},  {\em Phys. Rev. D}
  {\bf 86} (2012) 035007, [\href{http://arxiv.org/abs/1206.2590}{{\tt
  arXiv:1206.2590}}].

\bibitem{Dias:2011sq}
A.~G. Dias, C.~A. de~S.~Pires, and P.~S.~R. da~Silva, {\it {How the Inverse
  See-Saw Mechanism Can Reveal Itself Natural, Canonical and Independent of the
  Right-Handed Neutrino Mass}},  {\em Phys. Rev. D} {\bf 84} (2011) 053011,
  [\href{http://arxiv.org/abs/1107.0739}{{\tt arXiv:1107.0739}}].

\bibitem{CentellesChulia:2020dfh}
S.~Centelles~Chuli\'a, R.~Srivastava, and A.~Vicente, {\it {The inverse seesaw
  family: Dirac and Majorana}},  {\em JHEP} {\bf 03} (2021) 248,
  [\href{http://arxiv.org/abs/2011.06609}{{\tt arXiv:2011.06609}}].

\bibitem{Pinheiro:2021mps}
J.~a.~P. Pinheiro, C.~A. de~S.~Pires, F.~S. Queiroz, and Y.~S. Villamizar, {\it
  {Confronting the inverse seesaw mechanism with the recent muon g-2 result}},
  {\em Phys. Lett. B} {\bf 823} (2021) 136764,
  [\href{http://arxiv.org/abs/2107.01315}{{\tt arXiv:2107.01315}}].

\bibitem{Pongkitivanichkul:2019cvm}
C.~Pongkitivanichkul, N.~Thongyoi, and P.~Uttayarat, {\it {Inverse seesaw
  mechanism and portal dark matter}},  {\em Phys. Rev. D} {\bf 100} (2019),
  no.~3 035034, [\href{http://arxiv.org/abs/1905.13224}{{\tt
  arXiv:1905.13224}}].

\bibitem{Abada:2021yot}
A.~Abada, N.~Bernal, A.~E.~C. Hern\'andez, X.~Marcano, and G.~Piazza, {\it
  {Gauged inverse seesaw from dark matter}},  {\em Eur. Phys. J. C} {\bf 81}
  (2021), no.~8 758, [\href{http://arxiv.org/abs/2107.02803}{{\tt
  arXiv:2107.02803}}].

\bibitem{Nomura:2019xsb}
T.~Nomura, H.~Okada, and S.~Patra, {\it {An inverse seesaw model with $A_4$
  -modular symmetry}},  {\em Nucl. Phys. B} {\bf 967} (2021) 115395,
  [\href{http://arxiv.org/abs/1912.00379}{{\tt arXiv:1912.00379}}].

\bibitem{An:2011uq}
H.~An, P.~S.~B. Dev, Y.~Cai, and R.~N. Mohapatra, {\it {Sneutrino Dark Matter
  in Gauged Inverse Seesaw Models for Neutrinos}},  {\em Phys. Rev. Lett.} {\bf
  108} (2012) 081806, [\href{http://arxiv.org/abs/1110.1366}{{\tt
  arXiv:1110.1366}}].

\bibitem{Biswas:2018yus}
A.~Biswas, S.~Choubey, and S.~Khan, {\it {Inverse seesaw and dark matter in a
  gauged B-L extension with flavour symmetry}},  {\em JHEP} {\bf 08} (2018)
  062, [\href{http://arxiv.org/abs/1805.00568}{{\tt arXiv:1805.00568}}].

\bibitem{Esteban:2020cvm}
I.~Esteban, M.~C. Gonzalez-Garcia, M.~Maltoni, T.~Schwetz, and A.~Zhou, {\it
  {The fate of hints: updated global analysis of three-flavor neutrino
  oscillations}},  {\em JHEP} {\bf 09} (2020) 178,
  [\href{http://arxiv.org/abs/2007.14792}{{\tt arXiv:2007.14792}}].

\bibitem{Antusch:2005gp}
S.~Antusch, J.~Kersten, M.~Lindner, M.~Ratz, and M.~A. Schmidt, {\it {Running
  neutrino mass parameters in see-saw scenarios}},  {\em JHEP} {\bf 03} (2005)
  024, [\href{http://arxiv.org/abs/hep-ph/0501272}{{\tt hep-ph/0501272}}].

\bibitem{Huber:2004ka}
P.~Huber, M.~Lindner, and W.~Winter, {\it {Simulation of long-baseline neutrino
  oscillation experiments with GLoBES (General Long Baseline Experiment
  Simulator)}},  {\em Comput. Phys. Commun.} {\bf 167} (2005) 195,
  [\href{http://arxiv.org/abs/hep-ph/0407333}{{\tt hep-ph/0407333}}].

\bibitem{Huber:2007ji}
P.~Huber, J.~Kopp, M.~Lindner, M.~Rolinec, and W.~Winter, {\it {New features in
  the simulation of neutrino oscillation experiments with GLoBES 3.0: General
  Long Baseline Experiment Simulator}},  {\em Comput. Phys. Commun.} {\bf 177}
  (2007) 432--438, [\href{http://arxiv.org/abs/hep-ph/0701187}{{\tt
  hep-ph/0701187}}].

\bibitem{Morel:2020dww}
L.~Morel, Z.~Yao, P.~Clad\'e, and S.~Guellati-Kh\'elifa, {\it {Determination of
  the fine-structure constant with an accuracy of 81 parts per trillion}},
  {\em Nature} {\bf 588} (2020), no.~7836 61--65.

\bibitem{Davoudiasl:2018fbb}
H.~Davoudiasl and W.~J. Marciano, {\it {Tale of two anomalies}},  {\em Phys.
  Rev. D} {\bf 98} (2018), no.~7 075011,
  [\href{http://arxiv.org/abs/1806.10252}{{\tt arXiv:1806.10252}}].

\bibitem{Moore:1984eg}
S.~R. Moore, K.~Whisnant, and B.-L. Young, {\it {Second Order Corrections to
  the Muon Anomalous Magnetic Moment in Alternative Electroweak Models}},  {\em
  Phys. Rev. D} {\bf 31} (1985) 105.

\bibitem{Bardeen:1972vi}
W.~A. Bardeen, R.~Gastmans, and B.~E. Lautrup, {\it {Static quantities in
  Weinberg's model of weak and electromagnetic interactions}},  {\em Nucl.
  Phys. B} {\bf 46} (1972) 319--331.

\bibitem{KLOE-2:2014qxg}
{\bf KLOE-2} Collaboration, D.~Babusci et~al., {\it {Search for light vector
  boson production in $e^+e^- \rightarrow \mu^+ \mu^- \gamma$ interactions with
  the KLOE experiment}},  {\em Phys. Lett. B} {\bf 736} (2014) 459--464,
  [\href{http://arxiv.org/abs/1404.7772}{{\tt arXiv:1404.7772}}].

\bibitem{COHERENT:2017ipa}
{\bf COHERENT} Collaboration, D.~Akimov et~al., {\it {Observation of Coherent
  Elastic Neutrino-Nucleus Scattering}},  {\em Science} {\bf 357} (2017),
  no.~6356 1123--1126, [\href{http://arxiv.org/abs/1708.01294}{{\tt
  arXiv:1708.01294}}].

\bibitem{CCFR:1991lpl}
{\bf CCFR} Collaboration, S.~R. Mishra et~al., {\it {Neutrino tridents and W Z
  interference}},  {\em Phys. Rev. Lett.} {\bf 66} (1991) 3117--3120.

\bibitem{Aoyama:2020ynm}
T.~Aoyama et~al., {\it {The anomalous magnetic moment of the muon in the
  Standard Model}},  {\em Phys. Rept.} {\bf 887} (2020) 1--166,
  [\href{http://arxiv.org/abs/2006.04822}{{\tt arXiv:2006.04822}}].

\bibitem{Bodas:2021fsy}
A.~Bodas, R.~Coy, and S.~J.~D. King, {\it {Solving the electron and muon $g-2$
  anomalies in $Z'$ models}},  {\em Eur. Phys. J. C} {\bf 81} (2021), no.~12
  1065, [\href{http://arxiv.org/abs/2102.07781}{{\tt arXiv:2102.07781}}].

\bibitem{Kamada:2018zxi}
A.~Kamada, K.~Kaneta, K.~Yanagi, and H.-B. Yu, {\it {Self-interacting dark
  matter and muon $g-2$ in a gauged U$(1)_{L_{\mu} - L_{\tau}}$ model}},  {\em
  JHEP} {\bf 06} (2018) 117, [\href{http://arxiv.org/abs/1805.00651}{{\tt
  arXiv:1805.00651}}].

\bibitem{Araki:2014ona}
T.~Araki, F.~Kaneko, Y.~Konishi, T.~Ota, J.~Sato, and T.~Shimomura, {\it
  {Cosmic neutrino spectrum and the muon anomalous magnetic moment in the
  gauged $L_\mu-L_\tau$ model}},  {\em Phys. Rev. D} {\bf 91} (2015), no.~3
  037301, [\href{http://arxiv.org/abs/1409.4180}{{\tt arXiv:1409.4180}}].

\bibitem{Muong-2:2021ojo}
{\bf Muon g-2} Collaboration, B.~Abi et~al., {\it {Measurement of the Positive
  Muon Anomalous Magnetic Moment to 0.46 ppm}},  {\em Phys. Rev. Lett.} {\bf
  126} (2021), no.~14 141801, [\href{http://arxiv.org/abs/2104.03281}{{\tt
  arXiv:2104.03281}}].

\bibitem{Charity:2018sqj}
S.~Charity, {\em {Beam profile measurements using the straw tracking detectors
  at the Fermilab muon $g$ \ensuremath{-} 2 experiment, and a study of their
  sensitivity to a muon electric dipole moment}}.
\newblock PhD thesis, U. Liverpool (main), 2018.

\bibitem{Pilaftsis:1997jf}
A.~Pilaftsis, {\it {CP violation and baryogenesis due to heavy Majorana
  neutrinos}},  {\em Phys. Rev. D} {\bf 56} (1997) 5431--5451,
  [\href{http://arxiv.org/abs/hep-ph/9707235}{{\tt hep-ph/9707235}}].

\bibitem{Bambhaniya:2016rbb}
G.~Bambhaniya, P.~S. Bhupal~Dev, S.~Goswami, S.~Khan, and W.~Rodejohann, {\it
  {Naturalness, Vacuum Stability and Leptogenesis in the Minimal Seesaw
  Model}},  {\em Phys. Rev. D} {\bf 95} (2017), no.~9 095016,
  [\href{http://arxiv.org/abs/1611.03827}{{\tt arXiv:1611.03827}}].

\bibitem{Pilaftsis:2003gt}
A.~Pilaftsis and T.~E.~J. Underwood, {\it {Resonant leptogenesis}},  {\em Nucl.
  Phys. B} {\bf 692} (2004) 303--345,
  [\href{http://arxiv.org/abs/hep-ph/0309342}{{\tt hep-ph/0309342}}].

\bibitem{Abada:2018oly}
A.~Abada, G.~Arcadi, V.~Domcke, M.~Drewes, J.~Klaric, and M.~Lucente, {\it
  {Low-scale leptogenesis with three heavy neutrinos}},  {\em JHEP} {\bf 01}
  (2019) 164, [\href{http://arxiv.org/abs/1810.12463}{{\tt arXiv:1810.12463}}].

\bibitem{Buchmuller:2004nz}
W.~Buchmuller, P.~Di~Bari, and M.~Plumacher, {\it {Leptogenesis for
  pedestrians}},  {\em Annals Phys.} {\bf 315} (2005) 305--351,
  [\href{http://arxiv.org/abs/hep-ph/0401240}{{\tt hep-ph/0401240}}].

\bibitem{Aoki:2015owa}
M.~Aoki, N.~Haba, and R.~Takahashi, {\it {A model realizing inverse seesaw and
  resonant leptogenesis}},  {\em PTEP} {\bf 2015} (2015), no.~11 113B03,
  [\href{http://arxiv.org/abs/1506.06946}{{\tt arXiv:1506.06946}}].

\bibitem{Gu:2010xc}
P.-H. Gu and U.~Sarkar, {\it {Leptogenesis with Linear, Inverse or Double
  Seesaw}},  {\em Phys. Lett. B} {\bf 694} (2011) 226--232,
  [\href{http://arxiv.org/abs/1007.2323}{{\tt arXiv:1007.2323}}].

\bibitem{Sakharov:1967dj}
A.~D. Sakharov, {\it {Violation of CP Invariance, C asymmetry, and baryon
  asymmetry of the universe}},  {\em Pisma Zh. Eksp. Teor. Fiz.} {\bf 5} (1967)
  32--35.

\bibitem{Plumacher:1996kc}
M.~Plumacher, {\it {Baryogenesis and lepton number violation}},  {\em Z. Phys.
  C} {\bf 74} (1997) 549--559, [\href{http://arxiv.org/abs/hep-ph/9604229}{{\tt
  hep-ph/9604229}}].

\bibitem{Giudice:2003jh}
G.~F. Giudice, A.~Notari, M.~Raidal, A.~Riotto, and A.~Strumia, {\it {Towards a
  complete theory of thermal leptogenesis in the SM and MSSM}},  {\em Nucl.
  Phys. B} {\bf 685} (2004) 89--149,
  [\href{http://arxiv.org/abs/hep-ph/0310123}{{\tt hep-ph/0310123}}].

\bibitem{Strumia:2006qk}
A.~Strumia, {\it {Baryogenesis via leptogenesis}},  in {\em {Les Houches Summer
  School on Theoretical Physics: Session 84: Particle Physics Beyond the
  Standard Model}}, pp.~655--680, 8, 2006.
\newblock \href{http://arxiv.org/abs/hep-ph/0608347}{{\tt hep-ph/0608347}}.

\bibitem{Iso:2010mv}
S.~Iso, N.~Okada, and Y.~Orikasa, {\it {Resonant Leptogenesis in the Minimal
  B-L Extended Standard Model at TeV}},  {\em Phys. Rev. D} {\bf 83} (2011)
  093011, [\href{http://arxiv.org/abs/1011.4769}{{\tt arXiv:1011.4769}}].

\bibitem{Davidson:2008bu}
S.~Davidson, E.~Nardi, and Y.~Nir, {\it {Leptogenesis}},  {\em Phys. Rept.}
  {\bf 466} (2008) 105--177, [\href{http://arxiv.org/abs/0802.2962}{{\tt
  arXiv:0802.2962}}].

\bibitem{Planck:2015fie}
{\bf Planck} Collaboration, P.~A.~R. Ade et~al., {\it {Planck 2015 results.
  XIII. Cosmological parameters}},  {\em Astron. Astrophys.} {\bf 594} (2016)
  A13, [\href{http://arxiv.org/abs/1502.01589}{{\tt arXiv:1502.01589}}].

\bibitem{Cyburt:2015mya}
R.~H. Cyburt, B.~D. Fields, K.~A. Olive, and T.-H. Yeh, {\it {Big Bang
  Nucleosynthesis: 2015}},  {\em Rev. Mod. Phys.} {\bf 88} (2016) 015004,
  [\href{http://arxiv.org/abs/1505.01076}{{\tt arXiv:1505.01076}}].

\bibitem{Planck:2018vyg}
{\bf Planck} Collaboration, N.~Aghanim et~al., {\it {Planck 2018 results. VI.
  Cosmological parameters}},  {\em Astron. Astrophys.} {\bf 641} (2020) A6,
  [\href{http://arxiv.org/abs/1807.06209}{{\tt arXiv:1807.06209}}]. [Erratum:
  Astron.Astrophys. 652, C4 (2021)].

\end{thebibliography}\endgroup

\end{document}